\documentclass[aps,prb,twocolumn,floatfix,superscriptaddress,nofootinbib]{revtex4-2}
\usepackage{graphicx}
\usepackage{bbm}
\usepackage{amsmath}
\usepackage{amssymb}
\usepackage{amsfonts}
\usepackage{lipsum}
\usepackage[colorlinks=true,allcolors=blue]{hyperref}
\usepackage{ulem}
\usepackage{slashed}
\usepackage{tikz}
\usepackage[compat=1.1.0]{tikz-feynman}
\usetikzlibrary{decorations.markings}
\usepackage{tensor}
\usepackage{comment}
\usepackage{subfigure}
\usepackage{booktabs}

\usepackage{braket} 
\usepackage{tensor} 

\newcommand{\mycomment}[1]{}

\newcommand{\nn}{\nonumber}

\newcommand{\td}{\text{d}}
\newcommand{\sgn}{\text{sign}}

\def\emph{\textit}

\begin{document}

\begin{abstract}
We study the torsional Hall viscosity $\zeta_{\rm H}$ of Chern insulators at finite temperature and chemical potential. Our analysis considers the massive Dirac fermion and two deformations: a constant out-of-plane magnetic field and a momentum-dependent mass described by the single-block Bernevig--Hughes--Zhang (BHZ) Hamiltonian. For the massive Dirac theory, we show that temperature and chemical potential lead to UV-finite corrections and that a chemical potential outside the bulk gap modifies $\zeta_{\rm H}$ even at zero temperature. In a constant magnetic field, the response of massive Dirac fermions is dictated by the relativistic Landau-level spectrum, leading to a plateau structure at zero temperature that is smoothed at finite temperature. The jumps are proportional to the Dirac mass and inversely proportional to the squared magnetic length and the newly occupied Landau-level energy. Finally, we present a detailed study of the correction to $\zeta_{\rm H}$ in the single-block BHZ continuum model due to temperature and chemical potential. For parameters relevant to (Hg,Mn)Te quantum wells, the magnitude of the jump across a mass domain wall retains its vacuum enhancement due to the BHZ deformation as long as the chemical potential lies sufficiently deep inside the gap compared with the thermal energy. Our results extend the theory of torsional Hall viscosity to experimentally relevant conditions and may facilitate its future detection.
\end{abstract}

\title{Torsional Hall Viscosity of Massive Chern Insulators: \\ Finite Temperature and Chemical Potential
}

\author{Weizhen Jia}
\email{Corresponding author: weizhenjia@cuhk.edu.hk}
\affiliation{Department of Physics, The Chinese University of Hong Kong, Shatin, New Territories, Hong Kong, China}
\thanks{The first two authors contributed equally to this work.}
\affiliation{Institute for Theoretical Physics and Astrophysics, and W{\"u}rzburg-Dresden Cluster of Excellence on Complexity, Topology and Dynamics in Quantum Matter ctd.qmat, Julius-Maximilians-Universit{\"a}t W{\"u}rzburg, Am~Hubland, D-97074~W{\"u}rzburg, Germany}

\author{Ioannis~Matthaiakakis}
\affiliation{Mathematical Sciences and STAG Research Centre, University of Southampton, Highfield,
Southampton SO17 1BJ, United Kingdom}
\thanks{The first two authors contributed equally to this work.}

\author{David Rodr\'iguez Fern\'andez}
\affiliation{Departamento de matemática aplicada a las TIC, Universidad Politécnica de Madrid, Nikola Tesla, s/n, ES-28031 Madrid, Spain}

\author{Zhuo-Yu~Xian}
\affiliation{School of Physics and Astronomy, Sun Yat-Sen University, 2 Daxue Road, Zhuhai 519082, China}
\affiliation{Department of Physics, Freie Universit\"at Berlin, Arnimallee 14, DE-14195 Berlin, Germany}

\author{Johanna~Erdmenger}
\affiliation{Institute for Theoretical Physics and Astrophysics, and W{\"u}rzburg-Dresden Cluster of Excellence on Complexity, Topology and Dynamics in Quantum Matter ctd.qmat, Julius-Maximilians-Universit{\"a}t W{\"u}rzburg, Am~Hubland, D-97074~W{\"u}rzburg, Germany}

\author{Ewelina~M.~Hankiewicz}
\affiliation{Institute for Theoretical Physics and Astrophysics, and W{\"u}rzburg-Dresden Cluster of Excellence on Complexity, Topology and Dynamics in Quantum Matter ctd.qmat, Julius-Maximilians-Universit{\"a}t W{\"u}rzburg, Am~Hubland, D-97074~W{\"u}rzburg, Germany}
\affiliation{Department of Physics, University of California, Berkeley, CA 94720, USA}

\author{Ren\'e~Meyer}
\email{Corresponding author: rene.meyer@uni-wuerzburg.de}
\affiliation{Institute for Theoretical Physics and Astrophysics, and W{\"u}rzburg-Dresden Cluster of Excellence on Complexity, Topology and Dynamics in Quantum Matter ctd.qmat, Julius-Maximilians-Universit{\"a}t W{\"u}rzburg, Am~Hubland, D-97074~W{\"u}rzburg, Germany}
\affiliation{Shanghai Institute for Mathematics and Interdisciplinary Sciences (SIMIS),
Shanghai, 200433, China}
\maketitle

\section{Introduction}
\label{Sec:Intro}
Hall viscosity is a nondissipative transport coefficient characterizing the transverse response of momentum to geometric deformations in systems with broken parity or time-reversal symmetry \cite{Avron:1995fg,Hoyos_2014}. In two-dimensional gapped systems, particularly quantum Hall states, it provides a geometric probe with connections to orbital spin and electromagnetic response \cite{Read2009,ReadRezayi2011,Hoyos:2011ez,BradlynGoldsteinRead2012,Bradlyn:2014wla}. For relativistic fermions, geometric deformations can be described by coupling the theory to a background geometry with torsion, and the corresponding parity-odd response defines the torsional Hall viscosity $\zeta_{\rm H}$ \cite{Hughes2011,Hughes2013,Parrikar:2014usa}. This description is particularly natural in crystalline systems, where torsion is associated with lattice dislocations \cite{KATANAEV19921,DEJUAN2010625}. Torsional geometry has been further developed as a framework for studying topological phases, quantum anomalies, and the thermal Hall response \cite{Cho2015,Valle2015,Gromov:2017vir,Huang:2019haq,Huang:2019adx,Huang:2021luf}.

The massive Dirac fermion provides the simplest continuum model in which the torsional Hall viscosity can be studied analytically. Its absolute bulk value contains an ultraviolet-sensitive contribution, whereas differences between distinct gapped phases, in particular the jump across a mass domain wall, are well defined once a common renormalization prescription is chosen \cite{Hughes2011,Hughes2013,Parrikar:2014usa}. However, the relativistic Dirac model provides only a simplified description of the electronic structure of realistic topological materials. A natural extension is provided by the Bernevig--Hughes--Zhang (BHZ) Hamiltonian, originally introduced to describe HgTe quantum wells \cite{Bernevig_2006}. The full BHZ Hamiltonian consists of two blocks related by time reversal and describes the quantum spin Hall phase, which was subsequently observed experimentally in HgTe/(Hg,Cd)Te quantum wells \cite{Ko_nig_2007}. Closely related BHZ-type effective Hamiltonians also describe inverted InAs/GaSb quantum wells \cite{PhysRevLett.100.236601,PhysRevB.81.201301} and arise in ultrathin films of three-dimensional topological insulators such as Bi$_2$Se$_3$, where the low-energy theory contains massive Dirac fermions with quadratic momentum-dependent mass terms \cite{BiSeTe,Lu_2010}. These systems provide experimentally accessible settings in which the simple Dirac description is replaced by a momentum-dependent band structure.

For the Hall-type response considered here, time-reversal symmetry must be broken. It is therefore natural to focus on a single block of the BHZ Hamiltonian, which by itself describes a massive Chern insulator. A physical motivation for this reduction is provided by magnetically doped HgTe quantum wells. In (Hg,Mn)Te systems, exchange coupling to polarized Mn moments breaks time-reversal symmetry and modifies the two BHZ sectors differently, leading to quantum anomalous Hall (QAH) physics, with a weak magnetic field applied to induce the Mn polarization \cite{Novik2005,Liu2008QAH,Shamim2020}. Such quantum wells and related topological insulators have been realized experimentally, and their topological transport properties have been studied in detail \cite{Shamim2020,Shamim2022,Wang2024}. The single-block BHZ Hamiltonian therefore provides a simple effective description of the parity-odd sector relevant for the torsional response of a topological quantum well with broken time-reversal symmetry.

Motivated by this connection, in Ref.~\cite{matthaiakakis2025} we extended the torsional Hall viscosity of the massive Dirac fermion in two directions. We first considered a constant electromagnetic field and obtained the response nonperturbatively in the field strength. We then studied the quadratic momentum-dependent mass that appears in a single block of the BHZ Hamiltonian. Although this deformation does not alter the low-energy Dirac structure, it substantially modifies the torsional response. In particular, for parameters relevant to (Hg,Mn)Te quantum wells, the torsional Hall viscosity jump across a mass domain wall was found to be enhanced relative to the free massive Dirac result. This enhancement makes the momentum-dependent model particularly interesting from an experimental point of view.

The analysis of Ref.~\cite{matthaiakakis2025}, however, was restricted to zero temperature and zero chemical potential. These assumptions are natural for isolating the ground-state response but are idealizations from the perspective of experiment. Measurements are necessarily performed at finite temperature, while the chemical potential in semiconductor quantum wells and thin films can be varied through doping or electrostatic gating. The occupation of the states contributing to the response therefore changes with both $T$ and $\mu$, especially when the chemical potential approaches a bulk band edge or crosses a Landau level. Extending the calculation to finite temperature and chemical potential is thus an important step toward determining whether the vacuum torsional Hall viscosity, and in particular the enhanced response found for the single-block BHZ model, can persist under experimentally accessible conditions.

In this work, we carry out this extension for the free massive Dirac fermion, for the Dirac fermion in a constant magnetic field, and for the single-block BHZ deformation. For the free Dirac theory, the finite-$T$ and finite-$\mu$ contribution is ultraviolet finite and can be separated from the renormalized vacuum response. At zero temperature, the response remains unchanged as long as the chemical potential lies inside the bulk gap, while finite temperature smoothens the sharp onset associated with crossing a band edge. In a constant magnetic field, the thermal correction can be expressed in terms of the Fermi--Dirac occupation of the relativistic Landau levels. The zeroth Landau level gives an additional orientation-sensitive contribution at finite chemical potential, while tuning $\mu$ through the higher Landau levels produces a sequence of nonquantized steps in $\zeta_{\rm H}$ that are rounded at finite temperature.

We finally consider the single-block BHZ model at finite temperature and chemical potential. We derive an ultraviolet-finite expression for the thermal correction and obtain analytic descriptions both away from and near the band edge. A central quantity in our analysis is the torsional Hall viscosity jump across a mass domain wall. When the chemical potential lies sufficiently deep inside the bulk gap, the enhanced vacuum response of Ref.~\cite{matthaiakakis2025} is stable up to exponentially suppressed thermal corrections. As the chemical potential approaches the band edge, thermal and finite-density effects increasingly modify the response. Using parameters relevant to (Hg,Mn)Te quantum wells, we determine the regime in which the zero-temperature prediction remains a good approximation. Therefore, the finite-$T$ and finite-$\mu$ analysis brings the theoretical torsional response closer to the conditions under which it may be tested experimentally in the future.

The paper is organized as follows. In Sec.~\ref{sec:Review}, we review the zero-temperature torsional Hall viscosity and summarize the magnetic-field and single-block BHZ results of Ref.~\cite{matthaiakakis2025}. In Sec.~\ref{sec:ThermalState}, we derive the response of the massive Dirac fermion at finite temperature and chemical potential. In Sec.~\ref{sec:MagField}, we consider a constant magnetic field and express the response in terms of the Landau-level spectrum. In Sec.~\ref{sec:BHZThermal}, we study the BHZ deformation and the finite-temperature domain-wall response. We conclude in Sec.~\ref{sec:Conclusions} with a discussion of the implications for torsional transport in Chern insulators and its possible experimental observation.

\section{Brief review of the vacuum results}
\label{sec:Review}
In this section, we summarize the zero-temperature and zero-chemical-potential results that will serve as the starting point for the thermal analysis below. We first recall the coupling between Dirac fermions and torsion and the corresponding definition of the torsional Hall viscosity. We then review the vacuum results for the free massive Dirac fermion, the constant magnetic field deformation, and the BHZ deformation. Throughout this section we work in units $\hbar=v_{\rm F}=k_{\rm B}=1$ unless stated otherwise.

\subsection{Torsional Hall viscosity and the effective action}
\label{sec:ReviewDefinition}
We consider a massive Dirac fermion in $(2+1)$ dimensions coupled to a background dreibein $e^a{}_\mu$ and an independent, metric-compatible spin connection $\omega^{ab}{}_\mu$. The Lorentzian action is
\begin{align}
\label{FreeDirac}
S_\psi=\int \td^3x\,e\bigg(\frac{1}{2}[\bar\psi\gamma^\mu\nabla_\mu\psi-(\nabla_\mu\bar\psi)\gamma^\mu\psi]-m\bar\psi\psi\bigg)\,,
\end{align}
where $e=|\det(e^a{}_\mu)|$. In terms of the coframe $e^a=e^a{}_\mu \td x^\mu$, the torsion 2-form is
\begin{align}
T^a=\td e^a+\omega^a{}_b\wedge e^b\,.
\end{align}
The spin connection can be decomposed as
\begin{align}
\omega^a{}_b=\mathring{\omega}^a{}_b+K^a{}_b\,,
\end{align}
where $\mathring{\omega}^a{}_b$ is the Levi-Civita spin connection and $K^a{}_b$ is the contorsion 1-form. Since $\td e^a+\mathring{\omega}^a{}_b\wedge e^b=0$, one equivalently has
\begin{align}
\label{Eq:TorsionContorsion}
T^a=K^a{}_b\wedge e^b\,.
\end{align}
Decomposing the fermion action in terms of the contorsion shows that a Dirac fermion couples only to its completely antisymmetric part. In $(2+1)$ dimensions this coupling can be expressed through the torsion pseudoscalar
\begin{align}
\label{Eq:ReviewSigma}
\sigma=\varepsilon^{abc}K_{[abc]}
=\varepsilon^{\lambda\mu\nu}K_{[\lambda\mu\nu]}\,,
\end{align}
and hence the part of the fermion action linear in torsion is
\begin{align}
\label{Eq:ReviewTorsionCoupling}
S_K=\frac{1}{4}\int d^3x\,e\,\sigma\bar\psi\psi\,.
\end{align}
Integrating out the fermion generates the parity-odd torsional term
\begin{align}
\label{Eq:ReviewEffectiveAction}
S_{\rm eff}
=\frac{\zeta_{\rm H}}{2}\int e^a\wedge T_a
=-\frac{\zeta_{\rm H}}{2}\int d^3x\,e\,\sigma\,,
\end{align}
where we used \eqref{Eq:TorsionContorsion} in the second equality. The coefficient $\zeta_{\rm H}$ defines the torsional Hall viscosity.\footnote{Note that $\zeta_{\rm H}$ is defined with the coframe and spin connection treated as independent sources and should be distinguished from the Hall viscosity defined through the generalized Belinfante stress response \cite{Bradlyn:2014wla}.}

More generally, if the torsion-dependent fermion action takes the form
\begin{align}
S_\psi=\int d^3x\,e\,\bar\psi\left(G^{-1}+V\sigma\right)\psi\,,
\end{align}
then the contribution linear in $\sigma$ gives
\begin{align}
\label{Eq:ZetaTadpole}
\zeta_{\rm H}=2V\lim_{y\rightarrow x}{\rm tr}\,G(x;y)\,.
\end{align}
Here the trace is over spinor indices and $G$ is the torsion-free fermion propagator, which may include external fields or higher-derivative deformations such as the magnetic-field and single-block BHZ deformations considered below. For the minimally coupled Dirac theory in Eq.~\eqref{FreeDirac}, the torsion vertex is $V=1/4$.

Eq.~\eqref{Eq:ZetaTadpole} reduces the calculation of $\zeta_{\rm H}$ to a one-loop tadpole. Its coincidence limit is generally UV divergent and must be regularized and renormalized. Consequently, the absolute bulk value of $\zeta_{\rm H}$ can contain scheme-dependent contributions from local counterterms. The difference in $\zeta_{\rm H}$ between phases, and in particular the jump across a domain wall, is a universal observable in the sense that it is insensitive to common phase-independent finite counterterms after the torsion coupling and vacuum subtraction prescription have been fixed~\cite{Hughes2011,Hughes2013,Parrikar:2014usa,matthaiakakis2025}.

\subsection{Massive Dirac vacuum}
\label{sec:ReviewDirac}

In flat Euclidean space, the propagator of the massive Dirac fermion reads
\begin{align}
\label{Eq:ReviewDiracPropagator}
G_0(k)=-\frac{i\slashed{k}+m}{k^2+m^2}\,.
\end{align}
Only the mass term survives the spinor trace in Eq.~\eqref{Eq:ZetaTadpole}. The vacuum torsional Hall viscosity can therefore be written as\footnote{With the Euclidean propagator in Eq.~\eqref{Eq:ReviewDiracPropagator}, extracting $\zeta_{\rm H}$ from the Wick-rotated Lorentzian effective action introduces an additional minus sign relative to Eq.~\eqref{Eq:ZetaTadpole}.}
\begin{align}
\label{Eq:ReviewDiracTadpole}
\zeta_{\rm H}^{\rm vac}=\frac{m}{2}\int\frac{d^3k}{(2\pi)^3}\,\frac{{\rm tr}\,\mathbbm{1}}{k^2+m^2}\,.
\end{align}
After regularization, the tadpole contains a linear UV divergence together with a finite parity-odd contribution. In the convention used in Ref.~\cite{matthaiakakis2025}, its regulated form is
\begin{align}
\label{Eq:ReviewDiracBare}
\zeta_{\rm H}^{\rm vac}=-\frac{\sgn(m)m^2}{4\pi}+\frac{m\Lambda}{2\pi^2}+O(\Lambda^{-1})\,.
\end{align}
The linearly divergent term is removed by a local counterterm. The counterterm also fixes which side of the mass transition is identified as the trivial phase. Choosing the $m>0$ phase to be trivial gives
\begin{align}
\label{Eq:ReviewDiracPhases}
2\pi\zeta_{\rm H}^{\rm vac}=\begin{cases}
0\,, & m>0\,,\\
m^2\,, & m<0\,.
\end{cases}
\end{align}
A symmetric prescription instead assigns equal and opposite bulk values to the two signs of $m$. These prescriptions differ only by a local, phase-independent finite counterterm and therefore predict the same jump across a mass domain wall.

For a domain wall separating the phases with $m=+|m|$ and $m=-|m|$, we define the universal jump by
\begin{align}
\label{Eq:ReviewDiracJump}
\Delta\zeta_{\rm H}^{0}\equiv\zeta_{\rm H}^{\rm vac}(+|m|)-\zeta_{\rm H}^{\rm vac}(-|m|)=-\frac{m^2}{2\pi}\,.
\end{align}
Once the subtraction convention is fixed, the sign of $\Delta\zeta_{\rm H}$ encodes the direction of the interfacial torsional response. Note that unlike dissipative bulk and shear viscosities, Hall viscosity is nondissipative and is not constrained to be positive; its sign determines the direction of the transverse momentum response relative to the applied strain.
\subsection{Constant magnetic field}
\label{sec:ReviewMagnetic}

We next recall the vacuum result in the presence of a constant magnetic field $B$ perpendicular to the two-dimensional system. Introducing the covariant Dirac operator
\begin{align}
\slashed{D}=\gamma^\mu(\partial_\mu-ieA_\mu)\,,
\end{align}
the torsional Hall viscosity is determined by the heat kernel of
\begin{align}
H_B=-\slashed{D}^{\,2}+m^2\,.
\end{align}
At $T=\mu=0$, the contribution proportional to the Dirac operator vanishes after the symmetric frequency integral, and hence the vacuum torsional Hall viscosity is determined entirely by the mass term. The trace of the coincident heat kernel is
\begin{align}
\label{Eq:ReviewMagneticHeatKernel}
{\rm tr}\,K(s;H_B)=\frac{\sqrt{\pi}}{4\pi^2l_B^3\sqrt{\tilde s}}\,e^{-\tilde m^2\tilde s}\coth(\tilde s)\,,
\end{align}
where
\begin{align}
l_B=\frac{1}{\sqrt{|eB|}}\,,\qquad\tilde s=\frac{s}{l_B^2}\,,\qquad\tilde m=ml_B\,.
\end{align}
After $\zeta$-function regularization, one obtains the exact result
\begin{align}
\label{Eq:ReviewVacuumMagnetic}
\zeta_{\rm H}^{B,{\rm vac}}=\frac{\sqrt{2|eB|}\,m}{16\pi}\left[2\zeta\left(\frac{1}{2},\frac{m^2}{2|eB|}\right)-\sqrt{\frac{2|eB|}{m^2}}\right]\,.
\end{align}
Here $\zeta(s,a)$ denotes the Hurwitz zeta function. Eq.~\eqref{Eq:ReviewVacuumMagnetic} is nonperturbative in the ratio $|eB|/m^2$ and resums the weak-field expansion to all orders. Restoring $\hbar$ and $v_{\rm F}$, it is useful to introduce the magnetic field scale
\begin{align}
\label{Eq:ReviewBc}
B_c=\frac{m^2}{|e|\hbar v_{\rm F}^2}\,.
\end{align}
Here $m$ denotes the gap parameter in energy units. The exact result can then be written as
\begin{align}
\label{Eq:ReviewVacuumMagneticSI}
\zeta_{\rm H}^{B,{\rm vac}}={}&\frac{\sgn(m)m^2}{16\pi\hbar v_{\rm F}^2}\sqrt{\frac{2|B|}{B_c}}\nn\\
&\times\Bigg[2\zeta\bigg(\frac{1}{2},\frac{B_c}{2|B|}\bigg)-\sqrt{\frac{2|B|}{B_c}}\Bigg]\,.
\end{align}
The $\zeta$-function-regularized result in Eq.~\eqref{Eq:ReviewVacuumMagneticSI} has several notable properties. First, it is odd under $m\rightarrow-m$ and even under $B\rightarrow-B$. The even dependence on $B$ follows from gauge and Lorentz invariance, which allow the magnetic field to enter through $B^2$. Second, within this prescription, the magnitude of the response decreases with increasing $|B|$ up to $|B|\simeq5.03\,B_c$, where the response vanishes and then changes sign.

Eq.~\eqref{Eq:ReviewVacuumMagnetic} uses the symmetric $\zeta$-function prescription, for which the $B\rightarrow0$ limit is $-m|m|/(4\pi)$. To recover the convention of Eq.~\eqref{Eq:ReviewDiracPhases}, one adds the finite counterterm $C(m)=m^2/(4\pi)$. Since this counterterm is independent of $B$, $T$, and $\mu$, it does not affect the change induced by the magnetic field or the thermal and chemical potential corrections.

One can also derive the torsional Hall response in a constant electric field, which contains nonperturbative contributions associated with Schwinger pair production and acquires a nonzero imaginary part, reflecting the instability of the driven vacuum \cite{matthaiakakis2025}. Since a constant electric field does not define thermal equilibrium in the absence of energy relaxation, we restrict our attention here to equilibrium backgrounds and omit the electric field analysis.
\subsection{BHZ deformation}
\label{sec:ReviewBHZ}

The second deformation considered in this work is the momentum-dependent mass appearing in a single spin block of the BHZ model. Its Hamiltonian is
\begin{align}
\label{Eq:ReviewBHZHamiltonian}
H_{\rm BHZ}=\begin{pmatrix}m-bk_\perp^2 & a(k_x+ik_y)\\a(k_x-ik_y) & -m+bk_\perp^2\end{pmatrix}\,,
\end{align}
where $k_\perp^2=k_x^2+k_y^2$. The corresponding spectrum is
\begin{align}
\label{Eq:ReviewBHZDispersion}
E_k=\sqrt{a^2k_\perp^2+\left(m-bk_\perp^2\right)^2}\,.
\end{align}
The parameter $a=\hbar v_{\rm F}$ controls the momentum-dependent hybridization between the electron-like and hole-like sectors and sets the characteristic Dirac velocity. The parameter $b$ controls the quadratic band curvature and is related to an effective Newtonian mass through $b=1/(2M)$ in units with $\hbar=v_{\rm F}=1$. Except when restoring physical units, we set $a=1$ below. The corresponding continuum Lagrangian is
\begin{align}
\label{Eq:ReviewBHZLagrangian}
\mathcal L_{\rm BHZ}=\bar\psi\slashed{\partial}\psi-m\bar\psi\psi-b\partial_i\bar\psi\partial^i\psi\,,\qquad i=1,2\,.
\end{align}
The higher-derivative term proportional to $b$ breaks Lorentz symmetry down to spatial rotations. Although it is an irrelevant operator in the infrared, it is a dangerously irrelevant operator since it controls the large-momentum behavior of the tadpole. We treat $b$ as a spurion and adopt the minimal geometric coupling in which torsion continues to enter only through the Dirac operator. The torsion vertex therefore remains $V=1/4$.

With $a=1$, the bare vacuum torsional Hall viscosity is
\begin{align}
\label{Eq:ReviewBHZTadpole}
\zeta_{\rm H}^{\rm vac}(m,b)=\frac{1}{4\pi}\int_0^\infty dk_\perp k_\perp\frac{m-bk_\perp^2}{\sqrt{k_\perp^2+\left(m-bk_\perp^2\right)^2}}\,.
\end{align}
In contrast to the free massive Dirac theory, the BHZ tadpole contains quadratic and logarithmic UV divergences. This occurs since the momentum-dependent effective mass $m-bk_\perp^2$, rather than the Dirac mass $m$, controls the large-momentum behavior. Consequently, the limits $b\rightarrow0$ and $\Lambda\rightarrow\infty$ do not commute. A consistent renormalization must remove the quadratic and logarithmic divergences associated with the $b$-dominated UV behavior. Recovering the Dirac result as $b\rightarrow0$ requires an additional renormalization prescription, as described in Ref.~\cite{matthaiakakis2025}, rather than simply taking the limit of the fixed-$b$ result.

Using the Pauli--Villars prescription, the renormalized bulk result can be written compactly as follows \cite{matthaiakakis2025}:
\begin{widetext}
\begin{align}
\label{Eq:ReviewBHZVacuum}
\zeta_{\rm H}^{\rm vac}(m,b)=\frac{1}{4\pi}\bigg[\frac{m}{2b}\big(\sgn(m)+\sgn(b)\big)+\frac{\sgn(m)-\sgn(b)}{8b^2}\ln(1+4|bm|)+m^2\big(1-\sgn(b)\big)\bigg]\,.
\end{align}
\end{widetext}
For fixed $b$, changing the sign of $m$ takes the system between two distinct band phases. Unlike in a commonly used renormalization of the pure Dirac theory, however, the renormalized torsional Hall viscosity does not vanish in either phase. Moreover, its absolute bulk value depends on the finite part of the renormalization prescription. The condition $\zeta_{\rm H}\neq0$ therefore cannot by itself determine whether a phase is topologically nontrivial. The topological assignment is instead fixed by the topology of the occupied band, as reflected in the single-block BHZ Hall conductivity \cite{Lu_2010,Bottcher:2019rrz,Tutschku:2020drw},
\begin{align}
\sigma_{\rm H}=\frac{e^2}{4\pi}\left[\sgn(m)+\sgn(b)\right]\,.
\end{align}
It follows that the phases with $bm>0$ are topologically nontrivial, whereas those with $bm<0$ are topologically trivial. This classification is consistent with a common renormalization of the parity-odd gauge and torsional responses. Although finite local counterterms affect the absolute bulk value of $\zeta_{\rm H}$, they do not change the band topology. Common phase-independent counterterms also leave the domain-wall jump unchanged.

Consider an interface at which $b$ remains fixed while the Dirac mass $m$ changes sign. Using the same subtraction convention as in the Dirac theory, we define the corresponding vacuum jump by
\begin{align}
\label{Eq:ReviewBHZJumpDefinition}
\Delta\zeta_{\rm H}^{\rm vac}=\zeta_{\rm H}^{\rm vac}(+|m|,b)-\zeta_{\rm H}^{\rm vac}(-|m|,b)\,.
\end{align}
It is convenient to normalize it by the pure Dirac jump in Eq.~\eqref{Eq:ReviewDiracJump},
\begin{align}
\label{Eq:ReviewBHZVacuumRatio}
\mathcal R_0\equiv\frac{\Delta\zeta_{\rm H}^{\rm vac}}{\Delta\zeta_{\rm H}^{0}}=-\bigg(\frac{a^2}{2|bm|}
+
\frac{a^4}{8b^2m^2}
\ln\bigg(
1+\frac{4|bm|}{a^2}
\bigg)\bigg)\,.
\end{align}
Thus, the BHZ band curvature enhances the universal interface response relative to the relativistic Dirac limit in magnitude, but reverses the sign. For a set of parameters appropriate for a (Hg,Mn)Te quantum well, $|m|=25~{\rm meV}$, $b=-1075~{\rm meV\,nm^2}$ and $a=365~{\rm meV\,nm}$, one finds
\begin{align}
\label{Eq:ReviewHgMnTeRatio}
\mathcal R_0\simeq-4.30\,.
\end{align}
Thus, the magnitude of the torsional Hall viscosity jump is enhanced by a factor $4.30$, while its sign is reversed relative to the Dirac limit. The finite-temperature BHZ analysis in Sec.~\ref{sec:BHZThermal} will determine how this vacuum enhancement is modified by thermal occupation and by tuning the chemical potential through the bulk band edge.

The results reviewed above provide the three vacuum limits against which the thermal calculations will be checked: the free massive Dirac response, the nonperturbative magnetic-field response, and the universal BHZ domain-wall response. Since finite temperature and chemical potential do not alter the short-distance structure of the theory, the same vacuum counterterms will apply below. Consequently, the corrections due to temperature and chemical potential can be isolated by subtracting the vacuum results, and the resulting differences are UV finite.

\section{Torsional Hall viscosity at finite temperature and chemical potential}
\label{sec:ThermalState}

In this section, we extend our results for the torsional Hall viscosity to finite temperature and chemical potential, in view of applying them to more realistic experimental setups. We begin by explaining how the formalism of Sec.~\ref{sec:Review} must be modified to include finite-temperature and finite-chemical-potential effects.

Recall that we work in Euclidean signature. We can therefore place the system at a finite temperature $T=1/\beta$ by compactifying the Euclidean time direction into a circle with period $\beta$. In every case considered here, the $x^3$ coordinate is a cyclic coordinate and can be identified with the Wick-rotated time coordinate. The fermion fields obey antiperiodic boundary conditions around the thermal circle. We can thus translate the vacuum results to finite temperature by replacing the integral over the momentum in the $x^3$ direction with a sum over the fermionic Matsubara frequencies
\begin{align}
\omega_l=\frac{2\pi}{\beta}\left(l+\frac{1}{2}\right)\,,\qquad l\in\mathbb Z\,.
\end{align}
More precisely, we make the substitution
\begin{align}
\label{Eq:VacToTemp}
\int\frac{dk_3}{2\pi}\longrightarrow\frac{1}{\beta}\sum_{\omega_l}\,.
\end{align}
We can further introduce a nonzero chemical potential by treating $\mu$ as a Lagrange multiplier that fixes the average charge of the system. Practically, this amounts to shifting the Matsubara frequencies according to
\begin{align}
\label{Eq:ChemicalPotentialShift}
\omega_l\longrightarrow\omega_l+i\mu\,.
\end{align}
The final expressions in this section are even in $\mu$, as required by particle-hole symmetry. At $\mu=0$, thermally excited particles and holes can produce a nonzero thermal correction even though their net charge density vanishes.

Let us first apply this prescription to the simplest case, namely a free massive Dirac fermion with Lagrangian given by Eq.~\eqref{FreeDirac}. For the vacuum torsional Hall viscosity $\zeta_{\rm H}^{\rm vac}$, Eq.~\eqref{Eq:ZetaTadpole} gives
\begin{align}
\label{Eq:VacuumZeta}
\zeta_{\rm H}^{\rm vac}&=m\int\frac{d^3k}{(2\pi)^3}\frac{1}{k^2+m^2}\nn\\
&=m\int\frac{dk_3}{2\pi}\int\frac{d^2k_\perp}{(2\pi)^2}\frac{1}{k_\perp^2+k_3^2+m^2}\,,
\end{align}
where $k_\perp^2\equiv k_1^2+k_2^2$ denotes the squared momentum in the spatial plane, while $k_3$ is the momentum along the Wick-rotated time direction. The overall normalization in Eq.~\eqref{Eq:VacuumZeta} includes the trace over the two-component spinor indices and is fixed by matching the vacuum result reviewed in Sec.~\ref{sec:Review}.

An expression for $\zeta_{\rm H}$ at finite temperature and chemical potential can now be read off directly from Eq.~\eqref{Eq:VacuumZeta}:
\begin{align}
\label{Eq:ZetaThermalIntegral}
\zeta_{\rm H}(\beta,\mu)=\frac{m}{\beta}\sum_{\omega_l}\int\frac{d^2k_\perp}{(2\pi)^2}\frac{1}{k_\perp^2+(\omega_l+i\mu)^2+m^2}\nn\\
=\frac{m}{4\pi\beta}\sum_{\omega_l}\int_0^\infty\frac{dy}{y+(\omega_l+i\mu)^2+m^2}\,,
\end{align}
where in the second equality we introduced polar coordinates in the spatial plane and defined $y=k_\perp^2$.

The resulting integral is logarithmically divergent in spatial momentum and must be regularized. Introducing a spatial momentum cutoff $y\leqslant\Lambda^2$, we find
\begin{align}
\label{Eq:ZetaMats}
\zeta_{\rm H}(\beta,\mu;\Lambda)=\frac{m}{4\pi\beta}\sum_{\omega_l}\ln\bigg[\frac{(\omega_l+i\mu)^2+\Lambda^2+m^2}{(\omega_l+i\mu)^2+m^2}\bigg]\,.
\end{align}
The sum over Matsubara frequencies can be evaluated in closed form, as shown in Appendix~\ref{App:MatsSum}. Taking $\Lambda\rightarrow\infty$ gives
\begin{widetext}
\begin{align}
\label{Eq:ThermalZeta}
\zeta_{\rm H}(\beta,\mu;\Lambda)=\frac{m\Lambda}{4\pi}-\frac{m}{4\pi\beta}\ln\left[4\cosh\left(\frac{\beta(|m|-\mu)}{2}\right)\cosh\left(\frac{\beta(|m|+\mu)}{2}\right)\right]+O(\Lambda^{-1})\,.
\end{align}
\end{widetext}
The linear divergence in Eq.~\eqref{Eq:ThermalZeta} is independent of temperature and chemical potential, and hence the same vacuum counterterms apply.\footnote{Here the coefficient of the divergent term differs from that in Eq.~\eqref{Eq:ReviewDiracBare} since we use a spatial momentum cutoff instead of a three-dimensional cutoff. The finite vacuum renormalization condition is the same.} Denoting the finite counterterm fixed by the vacuum prescription of Sec.~\ref{sec:Review} by $C(m)$, the renormalized response can be written as
\begin{align}
\label{Eq:DiracThermalRenormalizedFinitePart}
\zeta_{\rm H}^{\rm ren}(\beta,\mu;m)
=C(m)-\frac{m}{4\pi\beta}\ln[2\cosh(\beta|m|)+2\cosh(\beta\mu)]\,.
\end{align}
At zero chemical potential, this reduces to
\begin{align}
\label{Eq:FiniteTZeta}
\zeta_{\rm H}^{\rm ren}(\beta,0;m)=C(m)-\frac{m}{2\pi\beta}\ln\left[2\cosh\left(\frac{\beta m}{2}\right)\right]\,.
\end{align}
The corresponding renormalized vacuum response is
\begin{align}
\label{Eq:DiracVacuumRenormalizedFinitePart}
\zeta_{\rm H}^{\rm vac,ren}(m)=C(m)-\frac{m|m|}{4\pi}\,.
\end{align}
Later on, the superscript ``ren'' will be suppressed when there is no ambiguity.
The correction due to temperature and chemical potential is the difference between these responses, in which the common counterterm cancels:
\begin{align}
\label{Eq:DiracThermalCorrectionDefinition}
\delta\zeta_{\rm H}(\beta,\mu;m)\equiv\zeta_{\rm H}(\beta,\mu;m)-\zeta_{\rm H}^{\rm vac}(m)\,.
\end{align}
The full response therefore separates into its vacuum value and a UV-finite correction,
\begin{align}
\label{Eq:DiracThermalDecomposition}
\zeta_{\rm H}(\beta,\mu;m)=\zeta_{\rm H}^{\rm vac}(m)+\delta\zeta_{\rm H}(\beta,\mu;m)\,.
\end{align}
Therefore, the correction is~\cite{Valle2015}
\begin{widetext}
\begin{align}
\label{Eq:DiracThermalCorrection}
\delta\zeta_{\rm H}(\beta,\mu;m)=-\frac{m}{4\pi\beta}\left[\ln\left(1+e^{-\beta(|m|-\mu)}\right)+\ln\left(1+e^{-\beta(|m|+\mu)}\right)\right]\,.
\end{align}
\end{widetext}
This form makes explicit that the thermal and chemical-potential correction is UV finite.

Using the vacuum renormalization prescription introduced in Sec.~\ref{sec:Review},
\begin{align}
\label{Eq:DiracVacuumRenormalized}
\zeta_{\rm H}^{\rm vac}(m)=\frac{m^2-m|m|}{4\pi}\,,
\end{align}
the fully renormalized result becomes
\begin{align}
\label{Eq:ThermalZetaRenormalized}
\zeta_{\rm H}(\beta,\mu;m)=\frac{m^2}{4\pi}-\frac{m}{4\pi\beta}\ln\left[2\cosh(\beta|m|)+2\cosh(\beta\mu)\right]\,.
\end{align}
The first term is the finite counterterm fixed by the vacuum renormalization condition and is independent of temperature and chemical potential.

We can use Eq.~\eqref{Eq:ThermalZetaRenormalized} to recover the zero-temperature result and extend it to nonzero chemical potential. In the limit $\beta\rightarrow\infty$, we have
\begin{align}
\lim_{\beta\rightarrow\infty}\frac{1}{\beta}\ln\left[2\cosh(\beta|m|)+2\cosh(\beta\mu)\right]=\max(|m|,|\mu|)\,,
\end{align}
and hence
\begin{align}
\label{Eq:ZetaZeroT}
\zeta_{\rm H}(T=0,\mu;m)=\frac{m^2}{4\pi}-\frac{m}{4\pi}\max(|m|,|\mu|)\,.
\end{align}
When the chemical potential lies inside the bulk gap, $|\mu|<|m|$, we recover the vacuum result:
\begin{align}
\label{Eq:ZetaInsideGap}
\zeta_{\rm H}(T=0,|\mu|<|m|;m)=\zeta_{\rm H}^{\rm vac}(m)\,.
\end{align}
Thus, at zero temperature the torsional Hall viscosity is exactly independent of the chemical potential as long as the Fermi level remains inside the gap. In terms of the finite correction,
\begin{align}
\label{Eq:DiracZeroTemperatureCorrection}
\delta\zeta_{\rm H}(T=0,\mu;m)=-\frac{m}{4\pi}\left(|\mu|-|m|\right)\Theta\left(|\mu|-|m|\right)\,.
\end{align}
Once $|\mu|>|m|$, a Fermi surface appears and the occupation of bulk states modifies the torsional Hall response. This should not be interpreted as a change in the topology of the underlying Dirac bands. In this regime, the system is no longer insulating, and the response acquires a finite-density contribution from the occupied bulk states.

At finite temperature, the sharp zero-temperature threshold is thermally rounded. When the chemical potential lies well inside the gap, $|m|-|\mu|\gg T$, the leading correction is exponentially suppressed:
\begin{align}
\label{Eq:DiracActivatedCorrection}
\delta\zeta_{\rm H}(\beta,\mu;m)\simeq-\frac{m}{4\pi\beta}\left[e^{-\beta(|m|-\mu)}+e^{-\beta(|m|+\mu)}\right]\,.
\end{align}
The vacuum response is therefore exponentially stable at low temperature, with the activation scale determined by the distance between the chemical potential and the nearest band edge.

\section{Magnetic field deformation}\label{sec:MagField}

In the present section, we extend the analysis of section \ref{sec:ThermalState} by including an external constant magnetic field $B$, alongside temperature and chemical potential, to establish how the torsional Hall viscosity is affected. 

At zero magnetic field, as seen from Eq.~\eqref{Eq:ReviewDiracPropagator}, the part of the propagator proportional to the Dirac operator vanishes after the spinor trace, and hence the torsional Hall viscosity is determined entirely by the mass term. In a magnetic field, however, the Dirac operator and the corresponding heat kernel have a nontrivial Landau-level structure, and the Dirac-operator contribution need not vanish at finite chemical potential. Therefore, we separate the response into two parts:
\begin{align}
\zeta_{\rm H}^{B}
={}&\zeta_{{\rm H},m}^{B}
+\zeta_{{\rm H},\slashed D}^{B}\,,
\end{align}
where
\begin{align}
\label{Eq:ZetaBCompleteTrace}
\zeta_{{\rm H},m}^{B}
&=\frac{1}{2}\lim_{y\rightarrow x}{\rm tr}
\left[mH_B^{-1}(x;y)\right]\,,\\
\label{Eq:ZetaBDirac}
\zeta_{{\rm H},\slashed D}^{B}
&=\frac{1}{2}\lim_{y\rightarrow x}{\rm tr}
\left[\slashed D_xH_B^{-1}(x;y)\right]\,.
\end{align}
In Sec.~\ref{sec:MassTrace} we evaluate the mass-trace contribution $\zeta_{{\rm H},m}^{B}$, and in Sec.~\ref{sec:ExceptionalCompletion} we evaluate the Dirac-operator contribution $\zeta_{{\rm H},\slashed D}^{B}$.

\subsection{Mass-trace contribution and weak-field representation}
\label{sec:MassTrace}

For the mass-trace contribution $\zeta_{{\rm H},m}^B$, we start from the vacuum heat-kernel representation reviewed in Sec.~\ref{sec:ReviewMagnetic}. Leaving the integral over the Euclidean frequency explicit, we write
\begin{equation}
\label{Eq:ZetaVacuumB}
\begin{aligned}
    \zeta_{\rm H}^{B,{\rm vac}} &=\frac{m}{4\pi }\int ^{\infty}_{-\infty}\frac{\td p}{2 \pi }\bigg[ \int ^{\infty }_{0}\td s\,{\rm e}^{-( \tilde{p}^{2}+\tilde{m}^{2}) s}\coth ( s) \bigg] 
    \\
    &=\frac{m}{4\pi }\int ^{\infty}_{-\infty}\frac{\td p}{2\pi }\bigg[ \zeta\Big(s,{\tilde{p}^{2}+ \tilde{m}^{2}\over2}\Big) -{1\over \tilde{p}^{2}+\tilde{m}^{2}}\bigg]_{s=1}\,,
\end{aligned}
\end{equation}
where $\tilde{p}=pl_B$ and $\tilde{m}=ml_B$ are the dimensionless momentum and mass, with magnetic length $l_B=1/\sqrt{|eB|}$. In the second equality, we used the analytic continuation of an integral representation of the Hurwitz zeta function $\zeta(s,a)$. Since $s=1$ is a singular point of $\zeta(s,a)$, we leave the $s\to1$ limit implicit for the moment and return to it below. At finite $T$, $\mu$, and $B$, the mass-trace contribution can be written as
\begin{equation}
\label{Eq:ZetaTMBSum1}
\begin{aligned}
   \zeta_{{\rm H},m}^B
   = {m\over 4\pi\beta}\sum_{\omega_l}\bigg[ \zeta \Big( s,
  {\tilde{\omega}^{2}_{l,\mu}+ \tilde{m}^{2}\over2}\Big) -{1\over\tilde{\omega}^{2}_{l,\mu}+\tilde{m}^{2}}\bigg]_{s=1}\,,
\end{aligned}
\end{equation}
where we defined the shifted Matsubara frequency $\omega_{l,\mu}=\omega_l+i\mu$, with $\omega_l=2\pi(l+1/2)/\beta$.

We now discuss how to take the limit $s\rightarrow1$ in Eq.~\eqref{Eq:ZetaTMBSum1}. We begin by noting that $\zeta(s,a)$ satisfies
\begin{equation}
\label{Eq:HurZetaLim}
    \lim_{s\rightarrow 1} \left[\zeta(s,a)-{1\over s-1}\right] = -\psi(a)\,,
\end{equation}
where $\psi(z)$ is the digamma function. We use the formal relation \eqref{Eq:HurZetaLim} to obtain a well-defined expression for $\zeta_{{\rm H},m}^B$. Suppose that $a$ is indexed by an integer $l\in\mathbb{Z}$ and sum Eq.~\eqref{Eq:HurZetaLim} over $l$. Formally interchanging the sum and the limit gives
\begin{equation}
     \lim_{s\rightarrow 1} \left[\sum_a\zeta(s,a)-{1\over s-1}\sum_{a}1\right] = -\sum_a\psi(a)\,.
\end{equation}
The second term in the brackets contains the divergent sum $\sum_a1$. We regularize it by writing
\begin{equation}
\label{Eq:RenormConst}
\begin{aligned}
   &\sum_{l=-\infty}^\infty 1 = \lim_{\eta\rightarrow0^+} \sum_{l=-\infty}^\infty e^{i2\pi(l+1/2)\eta}
   \\
   &=\lim_{\eta\rightarrow0^+}e^{i\pi\eta}\left[\sum_{l=-\infty}^{0}(e^{i2\pi\eta})^l +\sum_{l=1}^\infty (e^{i2\pi\eta})^l\right]
   \\
   & = \lim_{\eta\rightarrow0^+}e^{i\pi\eta}\left({e^{i2\pi\eta}\over -1 +e^{i2\pi\eta}}- {e^{i2\pi\eta}\over -1 +e^{i2\pi\eta}}\right) = 0\,,
\end{aligned}
\end{equation}
where in the third equality we used the analytic continuation of the geometric series. This prescription assigns zero to the regulated constant and will be used in the following. Thus, for our purposes, we may write
\begin{equation}
    \sum_{a}\zeta(s\rightarrow 1,a) = -\sum_a\psi(a)\,,
\end{equation}
and hence
\begin{equation}
\label{Eq:ZetaTMBSum2}
\begin{aligned}
   \zeta_{{\rm H},m}^B
   =-{m\over 4\pi\beta}\sum_{\omega_l}\big[\psi( \tilde{\omega}^{2}_{l,\mu}/2+ \tilde{m}^{2}/2)+( \tilde{\omega}^{2}_{l,\mu}+\tilde{m}^{2}) ^{-1}\big]\,.
\end{aligned}
\end{equation}
The remaining Matsubara sum is not straightforward to perform due to the appearance of the digamma function. Nevertheless, we can obtain a perturbative expansion of Eq.~\eqref{Eq:ZetaTMBSum2}. As $|z|\rightarrow\infty$, the digamma function behaves as
\begin{equation}
    \psi(z)\sim \ln z -{1\over 2z}- \sum_{k=1}^\infty {B_{2k}\over 2k z^{2k}}\,,
\end{equation}
where $B_{2k}$ are the Bernoulli numbers. In our case, $z\propto1/|B|$, and hence this gives the following perturbative expansion in $|B|$ for $\zeta_{{\rm H},m}^B$:
\begin{equation}
\label{Eq:ZetaTMB1}
\begin{aligned}
    \zeta_{{\rm H},m}^B
&\sim -{m\over 4\pi\beta}\sum_{\omega_l}\bigg[\ln(\tilde{\omega}_{l,\mu}^2+\tilde{m}^2) \\
&\qquad\qquad\qquad\,\,-\sum_{k= 1}^\infty{4^k B_{2k}\over 2k}{1\over (\tilde{\omega}_{l,\mu}^2+\tilde{m}^2)^{2k}}\bigg]
    \\
    &=-{m\over 4\pi\beta}\sum_{\omega_l}\bigg[\ln(\omega_{l,\mu}^2 + m^2)\\
    &\qquad\qquad\qquad\,\,-\sum_{k= 1}^\infty{4^k B_{2k}\over 2k}{|eB|^{2k}\over (\omega_{l,\mu}^2+m^2)^{2k}}\bigg]\,,
\end{aligned}
\end{equation}
where we have renormalized away the infinite constant $\sum_{\omega_l}\ln(l_B^2/2)$ using the prescription in Eq.~\eqref{Eq:RenormConst}.

Eq.~\eqref{Eq:ZetaTMB1} naturally separates into a $B=0$ contribution and terms containing all dependence on the magnetic field through the sum over $k$. The Matsubara sum can be performed term by term, after which the series in $k$ can also be formally resummed. To this end, we use the relation
\begin{equation}
    {1\over (\omega_{l,\mu}^2+m^2)^k}
    = {(-1)^{k-1}\over (k-1)!}
    \partial^{k}_{m^2}
    \ln\left(\omega_{l,\mu}^2+m^2\right)\,.
\end{equation}
Subtracting the $T=\mu=0$ expression with the same regulator and using the linearity of the differential operator gives
\begin{align}
\label{Eq:ZetaTBM}
        \delta\zeta_{{\rm H},m}^B(\beta,\mu;m)
        =m {\partial \over \partial \tilde{m}^2}
        \coth\bigg({\partial \over \partial \tilde{m}^2}\bigg)
        \bigg[\frac{\delta\zeta_{\rm H}(\beta,\mu;m)}{m}\bigg]\,.
\end{align}
Here we reintroduced the dimensionless mass $\tilde{m}=ml_B$ and formally summed the series over $k$. Eq.~\eqref{Eq:ZetaTBM} gives the finite-temperature and chemical-potential correction to the mass-trace contribution $\zeta_{{\rm H},m}^B$.

The differential expression in Eq.~\eqref{Eq:ZetaTBM} can be
evaluated directly. Eq.~\eqref{Eq:DiracThermalCorrection} gives
\begin{align}
\label{Eq:ZetaBDifferentialDerivative}
\frac{\partial}{\partial m^2}
\bigg[
\frac{\delta\zeta_{\rm H}(\beta,\mu;m)}{m}
\bigg]
=
\frac{[n_{\rm F}(|m|-\mu)+n_{\rm F}(|m|+\mu)]}{8\pi|m|}\,,
\end{align}
where we have the Fermi–Dirac distribution
\begin{align}
    n_{\rm F}(x) = \frac{1}{e^{\beta x}+1}\,.
    \label{Eq:BHZThermalFermiFunction}
\end{align}
Using the spectral Laplace representation derived in Appendix~\ref{App:Laplace}, the differential operator in
Eq.~\eqref{Eq:ZetaTBM} therefore acts as
\begin{align}
\label{Eq:ZetaBCorrectTranslation}
&\frac{\partial}{\partial\widetilde m^2}
\coth\left(
\frac{\partial}{\partial\widetilde m^2}
\right)
\left[
\frac{\delta\zeta_{\rm H}(\beta,\mu;m)}{m}
\right]\nn\\
=&-\frac{1}{8\pi l_B^2}
\sum_{n=0}^{\infty}\frac{g_n}{E_n}
\left[
n_{\rm F}(E_n-\mu)+n_{\rm F}(E_n+\mu)
\right]\,,
\end{align}
where $n=0,1,\ldots$, and
\begin{align}
E_n=\sqrt{m^2+2n/l_B^2}\,, \qquad g_n=2-\delta_{n0}\,.
\end{align}
It follows that the finite-temperature and chemical-potential correction to the mass-trace contribution is
\begin{align}
\label{Eq:ZetaBDifferentialLandauSum}
\delta\zeta_{{\rm H},m}^{B}(\beta,\mu;m)
=&-\frac{m}{8\pi l_B^2}
\sum_{n=0}^{\infty}\frac{g_n}{E_n}
\left[n_{\rm F}(E_n-\mu)\right.\nn\\
&\left.\hspace{25mm}+n_{\rm F}(E_n+\mu)\right]\,.
\end{align}
Thus, Eq.~\eqref{Eq:ZetaBDifferentialLandauSum} is the Landau-level resummation of the mass-trace calculation in this subsection. The shifts $m^2\to m^2+2n/l_B^2$ reproduce the squared Dirac Landau-level energies $E_n^2$.

\subsection{Dirac-operator contribution}
\label{sec:ExceptionalCompletion}

We now evaluate the Dirac-operator contribution $\zeta_{{\rm H},\slashed D}^{B}$ in Eq.~\eqref{Eq:ZetaBDirac}. It vanishes at zero temperature and chemical potential after the symmetric frequency integral, but need not vanish when the Matsubara frequencies are shifted by $\mu$. To evaluate it in the same Landau-level basis as the mass-trace contribution, consider the single-particle Dirac Hamiltonian in the convention of Eq.~\eqref{Eq:ReviewBHZHamiltonian} at $b=0$,
\begin{align}
\label{Eq:DiracLandauHamiltonian}
h_B=\pi_x\sigma_1-\pi_y\sigma_2+m\sigma_3\,,
\qquad
\pi_i=-i\partial_i-eA_i\,.
\end{align}
For $B\neq0$, the paired Landau levels have energies $\pm E_n$ for $n\geqslant1$, while the zeroth Landau level $n=0$ is unpaired and has energy $\varepsilon_0=-\sgn(eB)m$.\footnote{The formulas below assume $m\neq0$.}

Each Landau level has degeneracy $1/(2\pi l_B^2)$ per unit area. In this basis, the corresponding operators take the form
\begin{align}
\label{Eq:ZetaBDiracOperators}
\slashed D(\omega_{l,\mu})
&=\sigma_3\left(h_B-m\sigma_3-i\omega_{l,\mu}\right)\,,\nn\\
H_B(\omega_{l,\mu})
&=\omega_{l,\mu}^2+h_B^2\,.
\end{align}
For each paired level $n\geqslant1$, we have $h_B^2=E_n^2\mathbbm{1}$ in the corresponding two-dimensional Landau-level sector. In the oscillator basis with $\sigma_3=\text{diag}(1,-1)$, the spatial Dirac operator is off diagonal. Choosing the relative phase of the two spinor components so that the off-diagonal matrix element is real and positive, we obtain
\begin{align}
\label{Eq:ZetaBDiracLandauBlock}
\slashed D H_B^{-1}\big|_n=\frac{1}{\omega_{l,\mu}^2+E_n^2}
\begin{pmatrix}
-i\omega_{l,\mu} & \sqrt{2n}/l_B\\
-\sqrt{2n}/l_B & i\omega_{l,\mu}
\end{pmatrix}\,,
\end{align}
where the subscript $n$ on the left-hand side denotes restriction to this sector. Taking the spinor trace gives
\begin{align}
\label{Eq:ZetaBDiracPairedTrace}
{\rm tr}\big[\slashed D H_B^{-1}\big|_n\big]
=\frac{-i\omega_{l,\mu}+i\omega_{l,\mu}}
{\omega_{l,\mu}^2+E_n^2}=0\,.
\end{align}
Thus, the paired levels provide no contribution to the Dirac-operator trace at any shifted Matsubara frequency.

The zeroth Landau level has only the spinor component with $\sigma_3$ eigenvalue $-\sgn(eB)$, and the spatial operator $h_B-m\sigma_3$ annihilates its state. Eq.~\eqref{Eq:ZetaBDiracOperators} therefore gives
\begin{align}
\left\langle 0\middle|\slashed D H_B^{-1}\middle|0\right\rangle
=\frac{i\sgn(eB)\omega_{l,\mu}}{\omega_{l,\mu}^2+m^2}\,.
\end{align}
Including the degeneracy $1/(2\pi l_B^2)$, Eq.~\eqref{Eq:ZetaBDirac} gives
\begin{align}
\label{Eq:ZetaBDiracMatsubara}
\zeta_{{\rm H},\slashed D}^{B}
=\frac{eB}{4\pi\beta}
\sum_{\omega_l}
\frac{i\omega_{l,\mu}}{\omega_{l,\mu}^2+m^2}\,,
\end{align}
where we used $eB=\sgn(eB)/l_B^2$. At $T=\mu=0$, the corresponding frequency integral vanishes since its integrand is odd.

As in Sec.~\ref{sec:ThermalState}, we isolate the finite-temperature and chemical-potential correction by subtracting the result at $T=\mu=0$ with the same regulator. To perform the remaining Matsubara sum, we write
\begin{equation}
\frac{i\omega_{l,\mu}}{\omega_{l,\mu}^2+m^2}=-\frac{1}{2}\bigg[\frac{1}{i\omega_l-\mu-|m|}
+\frac{1}{i\omega_l-\mu+|m|}\bigg]\,,
\end{equation}
and pair opposite Matsubara frequencies in the symmetric prescription. Using the standard Matsubara sum, for any $x\in\mathbb R$ we have
\begin{align}
\frac{1}{\beta}\sum_{\omega_l}\frac{1}{i\omega_l-x}
&=-\frac{x}{\beta}\sum_{\omega_l}\frac{1}{\omega_l^2+x^2}\nn\\
&=-\frac{1}{2}\tanh\left(\frac{\beta x}{2}\right)=n_{\rm F}(x)-\frac{1}{2}\,.
\end{align}
Together with $n_{\rm F}(-x)=1-n_{\rm F}(x)$, this gives the finite-temperature and chemical-potential correction for $\zeta_{{\rm H},\slashed D}^{B}$:
\begin{align}
\label{Eq:ZetaBExceptionalCorrection}
\delta\zeta_{{\rm H},\slashed D}^{B}
=\frac{eB}{8\pi}
\left[n_{\rm F}(|m|-\mu)-n_{\rm F}(|m|+\mu)\right]\,.
\end{align}

\subsection{Complete response and physical interpretation}
Combining the mass-trace contribution in Eq.~\eqref{Eq:ZetaBDifferentialLandauSum} with the Dirac-operator contribution in Eq.~\eqref{Eq:ZetaBExceptionalCorrection}, we obtain the complete correction
\begin{align}
\label{Eq:ZetaBFiniteCorrection}
&\delta\zeta_{\rm H}^{B}(\beta,\mu;m)\nn\\
=&-\frac{m}{8\pi l_B^2}
\sum_{n=0}^{\infty}\frac{g_n}{E_n}
[n_{\rm F}(E_n-\mu)+n_{\rm F}(E_n+\mu)]\nn\\
&+\frac{eB}{8\pi}
[n_{\rm F}(|m|-\mu)-n_{\rm F}(|m|+\mu)]\,.
\end{align}
The complete response is consequently
\begin{align}
\label{Eq:ZetaBFullRenormalized}
\zeta_{\rm H}^{B}(\beta,\mu;m)
=\zeta_{\rm H}^{B,{\rm vac}}(m)+\frac{m^2}{4\pi}
+\delta\zeta_{\rm H}^{B}(\beta,\mu;m)\,,
\end{align}
where $\zeta_{\rm H}^{B,{\rm vac}}$ is the symmetric vacuum result in Eq.~\eqref{Eq:ReviewVacuumMagnetic} and the same finite counterterm as at $B=0$ has been added.

Eq.~\eqref{Eq:ZetaBFiniteCorrection} contains both the contribution that depends only on $|eB|$ and a contribution from the zeroth Landau level proportional to $eB$. The latter vanishes at $\mu=0$ and changes sign under either $B\rightarrow-B$ or $\mu\rightarrow-\mu$. Thus, for a single Dirac block at $\mu\neq0$, the response depends on the sign of the magnetic field through the zeroth Landau level. Consequently, the complete correction $\delta\zeta_{\rm H}^{B}$ is invariant under the simultaneous transformation $(B,\mu)\rightarrow(-B,-\mu)$, as required by charge conjugation. In particular, the vacuum response and the thermal response at $\mu=0$ are even in $B$. 

A thermal state or a finite-density state provides a preferred timelike vector $u^\mu$. The scalar $u_\lambda\widetilde F^\lambda$, with $\widetilde F^\lambda=\varepsilon^{\lambda\mu\nu}F_{\mu\nu}/2$, reduces to the signed magnetic field in the rest frame. An orientation-sensitive equilibrium term is therefore allowed once its coefficient is odd in $\mu$.

The zero-field limit provides a direct check. For a smooth function $F(E_n)$, we have
\begin{align}
\label{Eq:LandauSumContinuumLimit}
\frac{1}{l_B^2}\sum_{n=0}^{\infty}g_nF(E_n)
\underset{B\rightarrow0}{\longrightarrow}
\int_0^{\infty}d(k_\perp^2)\,
F\left(\sqrt{k_\perp^2+m^2}\right)\,.
\end{align}
The contribution from the zeroth Landau level vanishes linearly with $B$, and Eq.~\eqref{Eq:ZetaBFiniteCorrection} reduces to
\begin{align}
\label{Eq:ZetaBZeroFieldCheck}
\delta\zeta_{\rm H}^{B\rightarrow0}
=-\frac{m}{4\pi}\int_{|m|}^{\infty}dE\,
\left[n_{\rm F}(E-\mu)+n_{\rm F}(E+\mu)\right]
=\delta\zeta_{\rm H}\,,
\end{align}
in agreement with Eq.~\eqref{Eq:DiracThermalCorrection}.

At zero temperature, Eq.~\eqref{Eq:ZetaBFiniteCorrection} becomes
\begin{align}
\label{Eq:ZetaBZeroTemperature}
\delta\zeta_{\rm H}^{B}(T=0,\mu;m)
=&-\frac{m}{8\pi l_B^2}
\sum_{n=0}^{\infty}\frac{g_n}{E_n}
\Theta(|\mu|-E_n)\nn\\
&+\frac{eB}{8\pi}\sgn(\mu)
\Theta(|\mu|-|m|)\,,
\end{align}
\begin{figure}[t]
    \centering
    \includegraphics[scale=0.58]{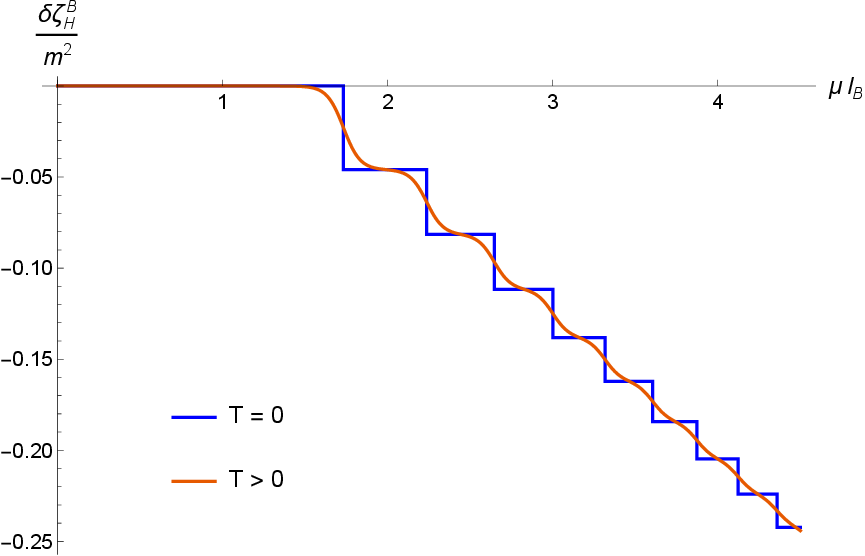}
    \caption{The finite-temperature and chemical-potential correction $\delta\zeta_{\rm H}^{B}$ as a function of $\mu l_B$ for $m l_B=1$ and $eB>0$, for which the magnetic length $l_B$ and the inverse gap scale $1/|m|$ are comparable. The orange curve illustrates how finite temperature rounds the steps as the chemical potential crosses successive Landau levels.}
    \label{fig:Prelim1}
\end{figure}
where $\Theta$ is the Heaviside step function. At a threshold we use the convention $\Theta(0)=1/2$. The response remains unchanged while the chemical potential lies in the spectral gap. At zero temperature, the paired-level steps occur at $|\mu|=E_n=\sqrt{m^2+2n/l_B^2}$ for $n\geqslant1$, while the threshold of the zeroth Landau level has magnitude $|\mu|=|m|$. Since the zeroth Landau level has signed energy $\varepsilon_0=-m\sgn(eB)$, this threshold occurs only on the corresponding side of the spectrum. In particular, for $m>0$ and $eB>0$, the zeroth Landau level lies at negative energy, whereas the first step at positive chemical potential occurs at $\mu=E_1$. Whenever $|\mu|$ crosses a paired Landau level $E_n$ with $n\geqslant1$, the response develops a new step. At the threshold of the zeroth Landau level $E_0=|m|$, the mass-trace and orientation-sensitive contributions may cancel or reinforce, depending on the signs of $m$, $eB$, and $\mu$. These steps are a spectral Landau-level effect rather than quantized torsional Hall plateaus, since the paired-level step heights contain the nonuniversal factor $m/E_n$. At finite temperature, each step is rounded over an energy scale of order $T$, smoothing the sharp edges of the zero-temperature plateaus. The resulting Landau-level steps and their thermal rounding are illustrated in Fig.~\ref{fig:Prelim1}.

For $|\mu|<|m|$ and $T\ll |m|-|\mu|$, all occupation factors are exponentially suppressed. Hence, $\delta\zeta_{\rm H}^{B}(\beta,\mu)$ vanishes exponentially, and the full response approaches the renormalized vacuum magnetic field result. By contrast, for $|\mu|\gg\max(|m|,T,l_B^{-1})$, many Landau levels are occupied and the sum approaches its continuum limit, giving $\delta\zeta_{\rm H}^{B}\sim-m|\mu|/(4\pi)$. This unbounded linear behavior is a formal property of the continuum Dirac model and should only be interpreted within the energy range where the low-energy Dirac description remains valid.

\section{BHZ deformation}
\label{sec:BHZThermal}

We now consider the BHZ deformation for a massive Dirac fermion in thermal equilibrium. The purpose of this section is to determine how the zero-temperature BHZ result is modified by finite temperature and chemical potential. 

As in the vacuum calculation, we set $a=1$ below. Physical units can be restored by replacing $b\rightarrow b/a^2$ and multiplying $\zeta_{\rm H}$ by $\hbar/a^2$, where $a=\hbar v_{\rm F}$. The two energy bands are
\begin{align}
E_\pm=\pm E_k\,,
\qquad
E_k
=
\sqrt{k_\perp^2+\left(m-bk_\perp^2\right)^2}\,,
\label{Eq:BHZThermalEnergy}
\end{align}
We will focus on the regime
\begin{align}
    bm<\frac{1}{2}\,.
    \label{Eq:BHZThermalSingleMinimum}
\end{align}
In this regime $E_k$ has a unique minimum at $k_\perp=0$, with band edges at $E=\pm|m|$. This is the regime relevant for the (Hg,Mn)Te quantum well in the quantum anomalous Hall experiments.

Following the finite-temperature prescription introduced in Sec.~\ref{sec:ThermalState}, the Euclidean frequency integral is replaced by a fermionic Matsubara sum, and the chemical potential is introduced by the shift $\omega_l\rightarrow \omega_l+i\mu$. Applying this prescription to the BHZ tadpole gives the formal expression
\begin{widetext}
\begin{align}
\zeta_{\rm H}(\beta,\mu;m,b)
=
\frac{1}{8\pi}
\int_0^\infty dk_\perp\,
\frac{k_\perp\left(m-bk_\perp^2\right)}{E_k}
\left[
\tanh\left(\frac{\beta(E_k-\mu)}{2}\right)
+
\tanh\left(\frac{\beta(E_k+\mu)}{2}\right)
\right]\,.
\label{Eq:BHZThermalIExact}
\end{align}
\end{widetext}
Like the vacuum BHZ tadpole, Eq.~\eqref{Eq:BHZThermalIExact} is UV divergent and is understood to be evaluated using the same regularization and counterterms as the vacuum response.
Since this expression is even in $\mu$, below we will consider $\mu$ to be positive. 

The vacuum BHZ calculation showed that the bare value of $\zeta_{\rm H}^{\rm vac}$ contains UV divergences. At finite temperature and chemical potential, the natural object is therefore the finite difference
\begin{align}
    \delta\zeta_{\rm H}(\beta,\mu;m,b)=\zeta_{\rm H}(\beta,\mu;m,b)-\zeta_{\rm H}^{\rm vac}(m,b)\,.
    \label{Eq:BHZThermalDeltaDefinition}
\end{align}
The two terms may be understood as either bare quantities evaluated with the same regulator or renormalized quantities evaluated using the same counterterms. This subtraction is convergent since finite temperature and chemical potential do not modify the UV structure of the BHZ tadpole. At large momentum the BHZ term dominates, with $E_k\sim |b|k_\perp^2$, and hence the divergent part is the same quadratic and logarithmic contribution already present in the vacuum calculation \cite{matthaiakakis2025}. After the Matsubara sum, subtracting the vacuum integrand leaves terms weighted by Fermi--Dirac factors, which are exponentially suppressed at large $k_\perp$ for fixed $\beta>0$ and finite $\mu$. Hence, $\delta\zeta_{\rm H}$ is UV finite and requires no additional regularization, while the same Pauli--Villars counterterms used at $T=\mu=0$ apply to the full response. We will not consider the formal high temperature expansion, since expanding the thermal factors before taking the UV limit is nonuniform and can obscure the vacuum power counting; moreover, the low temperature regime is the one relevant for experiments.

Using $\tanh(x/2)=1-2/(e^x+1)$, the finite correction \eqref{Eq:BHZThermalDeltaDefinition} can be written in terms of the Fermi--Dirac distribution $n_{\rm F}$ as
\begin{align}
\delta\zeta_{\rm H}(\beta,\mu;m,b)
={}&
-\frac{1}{4\pi}
\int_0^\infty dk_\perp\,
\frac{k_\perp\left(m-bk_\perp^2\right)}{E_k}
\nn\\
&\times
\left[
n_{\rm F}(E_k-\mu)
+
n_{\rm F}(E_k+\mu)
\right]\,.
\label{Eq:BHZThermalDeltaZetaExact}
\end{align}
Eq.~\eqref{Eq:BHZThermalDeltaZetaExact} is the main working formula of this section. Since it is UV finite, it can be evaluated numerically without additional regularization. In the following we will analyze this integral analytically in the low temperature regime.
For the numerical results in this section, we use a specific set of parameters $m=-25~{\rm meV}$, $b=-1075~{\rm meV\,nm^2}$, and $a=365~{\rm meV\,nm}$ valid for (Hg,Mn)Te quantum wells.

\subsection{Chemical potential correction at zero temperature}

At zero temperature, the Fermi–Dirac functions become step functions. If the chemical
potential lies inside the gap, i.e., $\mu<|m|$,
there is no Fermi surface and the correction vanishes exactly:
\begin{align}
    \delta\zeta_{\rm H}(T=0,\mu)=0\,,\qquad\mu<|m|\,.
    \label{Eq:BHZThermalNoCorrectionInsideGap}
\end{align}
Thus, at $T=0$, tuning the chemical potential does not change the torsional Hall viscosity as long as the Fermi level remains in the bulk gap.

For $\mu>|m|$, a Fermi surface appears. In the single-minimum regime $bm<1/2$, the Fermi momentum is determined by $E_{\rm F}=\mu$, giving
\begin{align}
    k_{\rm F}^2=\frac{-(1-2bm)+\sqrt{1-4bm+4b^2\mu^2}}{2b^2}\,.
    \label{Eq:BHZThermalKF}
\end{align}
The $\mu$-dependent part of the tadpole can then be evaluated analytically. Since Eq.~\eqref{Eq:BHZThermalNoCorrectionInsideGap}
shows that the zero-temperature response inside the gap is equal to the
vacuum response, the correction at zero temperature can be written as
\begin{align}
    \delta \zeta_{\rm H}(T=0,\mu)=
    \zeta_{\rm H}(T=0,\mu)-\zeta_{\rm H}(T=0,\mu_0)\,,
    \label{Eq:BHZThermalDeltaIDefZeroT}
\end{align}
with any $\mu_0<|m|$. For $\mu>|m|$ one finds\footnote{At $b<0$ and $bm=1/4$, the continuous limit gives $\delta\zeta_{\rm H}=\ln(\mu/|m|)/(16\pi b^2)+(\mu-|m|)/(8\pi b)$.}
\begin{align}
    \delta \zeta_{\rm H}(T=0,\mu)= &-\frac{1}{16\pi b^2}\ln\bigg[ \frac{2b\mu+\sqrt{1-4bm+4b^2\mu^2}}{2b|m|+1-2bm}\bigg]\nn\\
    &+\frac{1}{8\pi b}(\mu-|m|)\,.
    \label{Eq:BHZThermalDeltaIZeroT}
\end{align}
In the $b\rightarrow0$ limit, Eq.~\eqref{Eq:BHZThermalDeltaIZeroT} reduces to the free-Dirac zero-temperature correction in
Eq.~\eqref{Eq:DiracZeroTemperatureCorrection}.

\begin{figure}[h]
    \centering
    \includegraphics[scale=0.58]{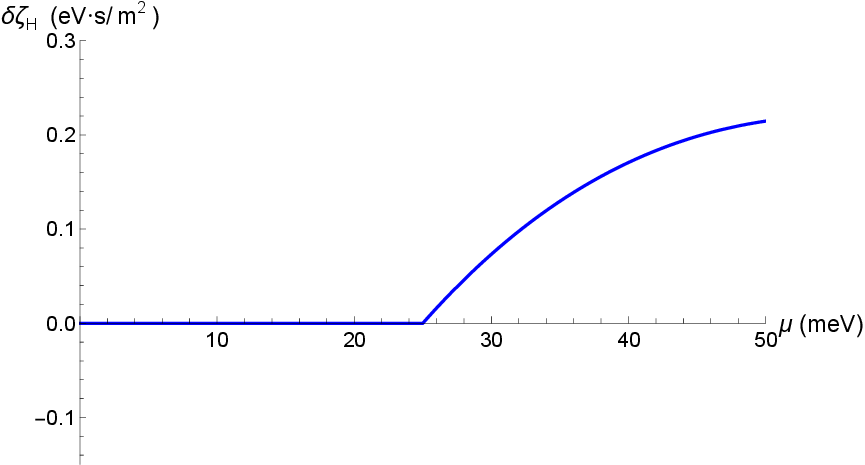}
    \caption{The zero-temperature chemical-potential correction $\delta\zeta_{\rm H}$ as a function of $\mu$. The correction vanishes inside the bulk gap and develops a kink when $\mu$ crosses the band edge.}
    \label{fig:BHZfinitemuzeroT}
\end{figure}

This correction vanishes continuously as $\mu\rightarrow |m|$ from above. Thus, the zero temperature response has a sharp kink at the band edge: below the gap there is no correction, while above the gap the Fermi surface contributes, see Fig.~\ref{fig:BHZfinitemuzeroT}.

\subsection{Low temperature expansion above the band edge}
\label{sec:BHZSommerfeld}

We next consider the regime in which the chemical potential lies above the band edge, i.e., $\mu>|m|$. At zero temperature this produces a Fermi surface, and the correction due to chemical potential is given by Eq.~\eqref{Eq:BHZThermalDeltaIZeroT}. At small but finite temperature, the natural analytic approximation is the Sommerfeld expansion. As we will see below, this expansion accurately describes the numerical result when $\mu$ is sufficiently far above the band edge, but it breaks down when $\mu$ approaches $|m|$.

\begin{figure}[h]
    \centering
        \includegraphics[scale=0.58]{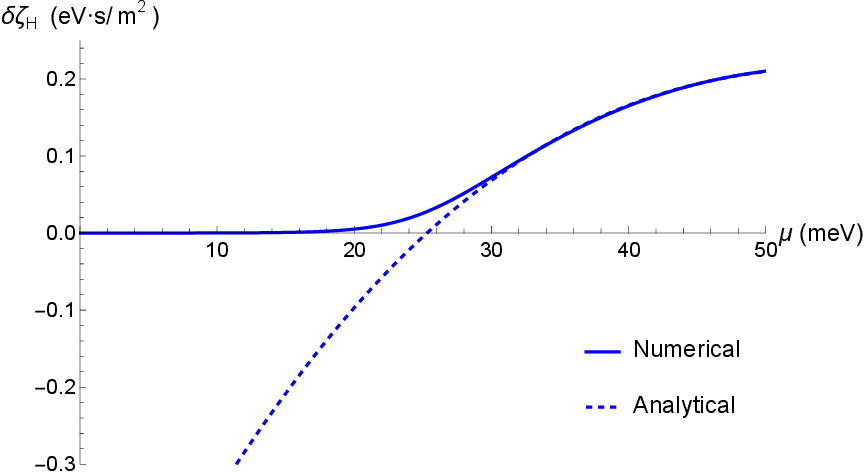}
    \caption{
    Comparison between the full numerical evaluation of $\delta\zeta_{\rm H}(\mu)$ and the Sommerfeld approximation at $T=30~{\rm K}$. The results agree when $\mu$ lies sufficiently above the band edge.}
    \label{fig:BHZSommerfeldValidity}
\end{figure}

In the single-minimum regime $bm<1/2$, the dispersion $E_k$ is monotonic for $k_\perp>0$. We may therefore change variables from $k_\perp$ to $E$ and define $\phi(E)$ by
\begin{align}
    dE\phi(E)=dk_\perp\frac{k_\perp\left(m-bk_\perp^2\right)}{E_k}\,.
    \label{Eq:BHZThermalPhiDef}
\end{align}
Solving for $\phi(E)$ we find
\begin{align}
    \phi(E)=\frac{1}{2b\sqrt{1-4bm+4b^2E^2}}-\frac{1}{2b}\,.
    \label{Eq:BHZThermalPhi}
\end{align}
The correction due to finite temperature contains the integral
\begin{align}
-\frac{1}{4\pi}
\int_{|m|}^{\infty}
dE\,\phi(E)n_{\rm F}(E-\mu)\,.
\label{Eq:BHZThermalSommerfeldIntegral}
\end{align}
where the contribution involving $n_{\rm F}(E+\mu)$ is exponentially suppressed at low temperature for positive $\mu$ above the edge of the conduction band. Applying the Sommerfeld expansion to Eq.~\eqref{Eq:BHZThermalSommerfeldIntegral} yields
\begin{align}
    \delta\zeta_{\rm H}(\beta,\mu)={}&\delta\zeta_{\rm H}(T=0,\mu)\nn\\
    &+\frac{\pi}{12\beta^2}\frac{b\mu}{\left(1-4bm+4b^2\mu^2\right)^{3/2}}+O(T^4)\,.
    \label{Eq:BHZThermalDeltaZetaSommerfeld}
\end{align}
The regime of validity of this expansion follows from the structure of the Fermi-Dirac distribution. At $T=0$, $n_{\rm F}(E-\mu)$ is a step function at $E=\mu$. At finite temperature, this step is smoothed over an energy interval of order $k_B T$, and hence the finite-temperature correction is controlled by energies satisfying $|E-\mu|\lesssim k_B T$. The Sommerfeld expansion assumes that the integrand is smooth throughout this interval and that the lower endpoint of the integral, $E=|m|$, is not inside it. Hence, the approximation is controlled when
\begin{align}
    \mu-|m|\gg k_B T\,.
    \label{Eq:BHZThermalSommerfeldValidity}
\end{align}
This explains the comparison with the numerical evaluation of Eq.~\eqref{Eq:BHZThermalDeltaZetaExact} shown in Fig.~\ref{fig:BHZSommerfeldValidity}: the Sommerfeld approximation agrees well when $\mu$ is sufficiently above the band edge, but fails as $\mu\rightarrow |m|$, where the lower endpoint begins to affect the finite $T$ crossover. 

For a fixed chemical potential above the band edge, the same criterion also gives a useful estimate of the temperature range where the Sommerfeld result should remain accurate. The endpoint correction omitted by the Sommerfeld expansion is controlled by $\exp[-(\mu-|m|)/(k_BT)]$. Thus, the validity scale $T_{\rm S}$ for the Sommerfeld expansion is given by $k_BT_{\rm S}=(\mu-|m|)/\ln(1/\epsilon)$, with $\epsilon$ a chosen error tolerance. For example, taking $|m|=25~{\rm meV}$ and $\mu=45~{\rm meV}$, a $5\%$ criterion gives $T_{\rm S}\simeq77~{\rm K}$, while a $1\%$ criterion gives $T_{\rm S}\simeq50~{\rm K}$ (see Appendix~\ref{app:BHZValidity} for details). Equivalently, at $T=40~{\rm K}$ the endpoint factor is already only $\exp[-20~{\rm meV}/(k_BT)]\simeq3\times10^{-3}$. This explains why the Sommerfeld approximation is expected to be very accurate through the low temperature regime relevant for QAH experiments, provided $\mu$ is sufficiently above the band edge.

In the Dirac limit $b\to0$, the energy space integrand becomes constant, and hence all algebraic Sommerfeld corrections vanish. Thus, for $\mu-|m|\gg k_B T$, the finite temperature Dirac result agrees with the $T=0$ result up to exponentially small endpoint corrections of order $\exp[-(\mu-|m|)/(k_B T)]$. The nonzero $T^2$ term in Eq.~\eqref{Eq:BHZThermalDeltaZetaSommerfeld} is therefore a genuine effect of the BHZ deformation.

\subsection{Thermal rounding of the band edge}

The Sommerfeld expansion fails when $\mu$ approaches $|m|$. In this regime the integral is dominated by states near the bottom of the band, and the sharp zero-temperature kink is thermally rounded. This is the physically important regime for understanding why $\zeta_{\rm H}$ can deviate from its zero temperature insulating value even before $\mu$ reaches the band edge.

Using Eq.~\eqref{Eq:BHZThermalPhi}, the finite correction may be written as
\begin{align}
    \delta \zeta_{\rm H}(\beta,\mu)=-\frac{1}{4\pi}\int_{|m|}^{\infty} dE\phi(E)[n_{\rm F}(E-\mu)+n_{\rm F}(E+\mu)]\,.
    \label{Eq:BHZThermalDeltaIEnergy}
\end{align}
Expanding $\phi(E)$ around $E=|m|$ gives the band-edge expansion
\begin{widetext}
\begin{align}
    \delta \zeta_{\rm H}(\beta,\mu)\simeq{}&-
    \frac{m}{4\pi(1-2bm)\beta}
    \left[\ln\left(1+e^{-\beta(|m|-\mu)}\right)+\ln\left(1+e^{-\beta(|m|+\mu)}\right)\right]
    \nonumber\\
    &-\frac{b|m|}{2\pi(1-2bm)^3\beta^2}\left[{\rm Li}_2\left(-e^{-\beta(|m|-\mu)}\right)+{\rm Li}_2\left(-e^{-\beta(|m|+\mu)}\right)\right]
    \nonumber\\
    &+\frac{b(-1+4bm+8b^2m^2)}{2\pi(1-2bm)^5\beta^3}
    \left[{\rm Li}_3\left(-e^{-\beta(|m|-\mu)}\right)+{\rm Li}_3\left(-e^{-\beta(|m|+\mu)}\right)\right]+\cdots\,,
    \label{Eq:BHZThermalDeltaIBandEdge}
\end{align}
\end{widetext}
where the polylogarithm function ${\rm Li}_n$ appears at $n$-th order in the expansion. As a check, in the Dirac limit $b\rightarrow0$, the first line of Eq.~\eqref{Eq:BHZThermalDeltaIBandEdge} reduces precisely to the free-Dirac thermal and chemical potential correction in Eq.~\eqref{Eq:DiracThermalCorrection}.

The leading term in Eq.~\eqref{Eq:BHZThermalDeltaIBandEdge} makes the physical crossover at the band edge transparent. For positive $\mu$ near the edge of the conduction band, the second logarithm is exponentially smaller than the first one, as it is controlled by $|m|+\mu$. The dominant crossover is therefore governed by $\ln(1+e^{-\beta(|m|-\mu)})/\beta$, which gives the smoothing of the kink at zero temperature. Indeed, in the zero temperature limit we have
\begin{align}
    \lim_{\beta\to\infty}
    \frac{1}{\beta}\ln\big(1+e^{-\beta(|m|-\mu)}\big)
    =(\mu-|m|)\Theta(\mu-|m|)\,.
    \label{Eq:BHZThermalSoftplusLimit}
\end{align}
Thus, the $T=0$ statement that $\zeta_{\rm H}$ is unchanged for $\mu<|m|$ is replaced at finite temperature by a smooth crossover of width
$O(k_BT)$.

\begin{figure}[h]
    \centering
    \includegraphics[scale=0.6]{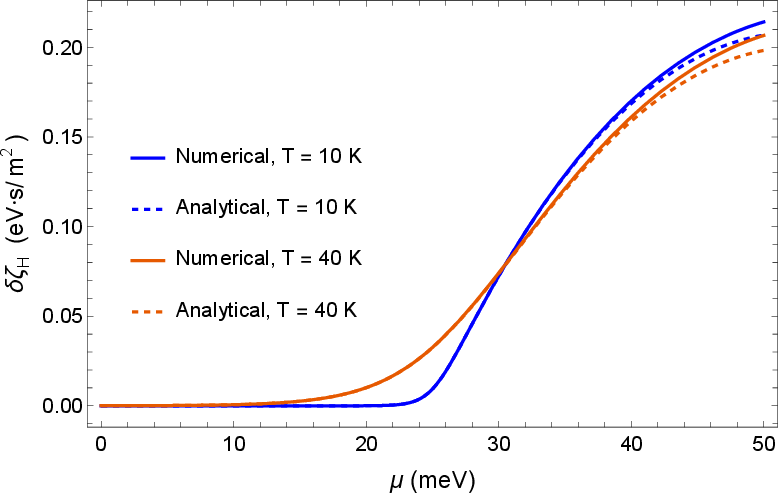}
    \caption{Numerical and analytical results for $\delta\zeta_H$ as a function of $\mu$ at fixed temperatures. The agreement is excellent when $\mu$ is not much greater than the band edge.}
    \label{fig:BHZbandedge}
\end{figure}

To estimate the chemical potential at which $\zeta_{\rm H}$ begins to deviate from its vacuum value, a useful quantitative diagnostic is the slope of the leading correction. Near the band edge we have
\begin{align}
    \frac{\partial\delta\zeta_{\rm H}}{\partial\mu}
    \simeq
    -\frac{m}{4\pi(1-2bm)}
    \frac{1}{1+e^{\beta(|m|-\mu)}}\,.
    \label{Eq:BHZThermalSlopeLogistic}
\end{align}
The sharp step at $T=0$ in slope is therefore rounded into a Fermi-Dirac function. If an onset value $\mu_{\rm on}$ is defined as the point where this slope
reaches a fraction $p$ of its asymptotic value, then
\begin{align}
    p=\frac{1}{1+e^{\beta(|m|-\mu_{\rm on})}}\,,
    \label{Eq:BHZThermalOnsetFraction}
\end{align}
from which we obtain
\begin{align}
    \mu_{\rm on}=|m|-k_BT\ln\bigg(\frac{1-p}{p}\bigg)\,.
    \label{Eq:BHZThermalOnsetMu}
\end{align}
If we set the visibility criterion $p=5\%$, then approximately we have 
\begin{align}
    \mu_{\rm on}=|m|-3k_BT\,.
    \label{Eq:BHZThermalOnsetMu5}
\end{align}
For $T=10$~K and $40$~K, we have $\mu_{\rm on}=22.4$~meV and $14.7$~meV, respectively. As can be seen in Fig.~\ref{fig:BHZbandedge}, the curves indeed begin to rise near these values of $\mu$.

The validity of the band edge expansion can be estimated analytically by comparing the first omitted term with the retained terms. For the (Hg,Mn)Te parameters and $\mu=30~{\rm meV}$ (5 meV above the band edge), this estimate shows that the approximation through ${\rm Li}_2$ begins to deviate at temperatures of a few tens of Kelvin, while including the ${\rm Li}_3$ term extends the useful range to temperatures of order $100~{\rm K}$. The details of this estimate are given in Appendix~\ref{app:BHZValidity}. Thus, the band edge expansion is well controlled in the low temperature regime relevant for QAH experiments, provided $\mu$ is not far above the band edge.

\subsection{Finite temperature at zero chemical potential}
\label{subsec:BHZThermalMuZero}

The band-edge expansion also gives a useful approximation at zero chemical potential. Setting $\mu=0$ in Eq.~\eqref{Eq:BHZThermalDeltaIBandEdge}, the particle and hole contributions become equal, and we obtain
\begin{align}
    \delta \zeta_{\rm H}(\beta,0)
    &\simeq-\frac{m}{2\pi(1-2bm)\beta}\ln\big(1+e^{-\beta |m|}\big)
    \nonumber\\
    &\quad-\frac{b|m|}{\pi(1-2bm)^3\beta^2}{\rm Li}_2\big(-e^{-\beta |m|}\big)\nonumber\\
    &+\frac{b(-1+4bm+8b^2m^2)}{\pi(1-2bm)^5\beta^3}{\rm Li}_3\big(-e^{-\beta |m|}\big)+\cdots \,.
    \label{Eq:BHZThermalDeltaIMuZero}
\end{align}
The physical content of Eq.~\eqref{Eq:BHZThermalDeltaIMuZero} is simple. At $T=0$ and $\mu=0$, there are no thermally excited carriers and the correction vanishes. At finite temperature, particle-hole excitations across the gap produce an exponentially small correction. For $\beta |m|\gg 1$, the leading term behaves as
\begin{align}
    \delta \zeta_{\rm H}(\beta,0)
    \simeq-\frac{m}{2\pi(1-2bm)\beta}e^{-\beta |m|}+O\bigg(\frac{e^{-\beta |m|}}{\beta^2}\bigg)\,.
    \label{Eq:BHZThermalDeltaIMuZeroLeading}
\end{align}
Thus, at charge neutrality the BHZ torsional Hall viscosity is exponentially stable at low temperature, with the activation scale set by the energy $|m|$ from the chemical potential to the band edge.

\subsection{Domain wall jump and physical interpretations}
\label{subsec:BHZThermalDomainWall}

The most robust observable of the torsional Hall
viscosity is the jump of $\zeta_{\rm H}$ across a domain wall where the Dirac mass changes sign, as reviewed in Sec.~\ref{sec:ReviewBHZ}. We use the same vacuum prescription and require both sides to lie in the single-minimum regime, i.e., $|bm|<1/2$, or $|bm|<a^2/2$ when $a$ is restored. At finite temperature and chemical potential, we decompose the jump across the domain wall as
\begin{align}
    \Delta\zeta_{\rm H}(\beta,\mu)=\Delta\zeta_{\rm H}^{\rm vac}+\Delta\delta\zeta_{\rm H}(\beta,\mu)\,,
    \label{Eq:BHZThermalJumpDecomposition}
\end{align}
where $\Delta\zeta_{\rm H}^{\rm vac}$ has been given in \eqref{Eq:ReviewBHZJumpDefinition}, and 
\begin{align}
    \Delta\delta\zeta_{\rm H}(\beta,\mu)=\delta\zeta_{\rm H}(\beta,\mu;+|m|,b)-\delta\zeta_{\rm H}(\beta,\mu;-|m|,b)\,.
    \label{Eq:BHZThermalDeltaJumpDefinition}
\end{align}
The thermal correction is finite without additional regularization and independent of the vacuum counterterms. The vacuum jump is evaluated using the prescription of Sec.~\ref{sec:ReviewBHZ}.

Using Eq.~\eqref{Eq:BHZThermalDeltaZetaExact}, the finite temperature correction to the domain wall jump can be written in the manifestly UV finite form
\begin{widetext}
\begin{align}
    \Delta\delta\zeta_{\rm H}(\beta,\mu)= -\frac{1}{8\pi b}\int_{|m|}^{\infty} dE\bigg[\frac{1}{\sqrt{1-4b|m|+4b^2E^2}}-\frac{1}{\sqrt{1+4b|m|+4b^2E^2}}\bigg]
    \left[n_{\rm F}(E-|\mu|)+n_{\rm F}(E+|\mu|)\right]\,.
    \label{Eq:BHZThermalExactJumpCorrection}
\end{align}
\end{widetext} 
 This integral is one of the main results of this section: for given $T$ and $\mu$, one can directly obtain numerically a prediction for how the domain wall response is modified in thermal equilibrium. Analytically, below we evaluate this integral in the same limits considered above for the bulk $\delta\zeta_{\rm H}$, which makes the physical content of the domain wall correction transparent.

First, at zero temperature, the correction vanishes if the chemical potential lies inside the gap:
\begin{align}
    \Delta\delta\zeta_{\rm H}(T=0,|\mu|<|m|)=0\,.
    \label{Eq:BHZThermalJumpInsideGap}
\end{align}
Thus, the vacuum domain wall response is exactly unchanged at $T=0$ as long as the Fermi level remains in the bulk gap. Once $|\mu|>|m|$, a Fermi surface appears and the jump receives a correction. In the single minimum regime
$|bm|<1/2$, this correction is\footnote{the expression at $b|m|=-1/4$ is understood by continuity.}
\begin{widetext}
\begin{align}
    \Delta\delta\zeta_{\rm H}(T=0,|\mu|)=\frac{1}{16\pi b^2}
    \ln\bigg[\frac{2b|\mu|+\sqrt{1+4b|m|+4b^2\mu^2}}{\left(1+4b|m|\right)\big(2b|\mu|+\sqrt{1-4b|m|+4b^2\mu^2}\big)}\bigg]\,,\qquad
    |\mu|>|m|\,.
    \label{Eq:BHZThermalJumpZeroTFiniteMu}
\end{align}
\end{widetext}
This expression vanishes continuously as $|\mu|\rightarrow |m|$.

At finite temperature the sharp zero temperature threshold is rounded. Near the band edge, the finite temperature correction to the domain wall jump can be obtained by applying the band edge expansion to the two sides of the wall. For convenience, we define
\begin{align}
    L_1(\beta,\mu)=\ln\big(1+e^{-\beta(|m|-|\mu|)}\big) +\ln\big(1+e^{-\beta(|m|+|\mu|)}\big)\,,
    \label{Eq:BHZThermalJumpL1}
\end{align}
and for $n\geqslant 2$,
\begin{align}
    L_n(\beta,\mu)={\rm Li}_n\big(-e^{-\beta(|m|-|\mu|)}\big)+{\rm Li}_n\big(-e^{-\beta(|m|+|\mu|)}\big)\,.
    \label{Eq:BHZThermalJumpLn}
\end{align}
Keeping terms up to ${\rm Li}_3$, the band edge expansion of the finite temperature correction to the domain wall jump is
\begin{widetext}
\begin{align}
    \Delta\delta\zeta_{\rm H}^{\rm BE}(\beta,\mu)
    \simeq-\frac{1}{2\pi}\bigg[&\frac{|m|}{\beta(1-4b^2m^2)}L_1(\beta,\mu)+\frac{4b^2m^2(3+4b^2m^2)}{\beta^2(1-4b^2m^2)^3}L_2(\beta,\mu)\nn\\
    &+\frac{4b^2|m|\left(3-80b^2m^2-464b^4m^4-128b^6m^6\right)}{\beta^3(1-4b^2m^2)^5}L_3(\beta,\mu)\bigg]\,.
    \label{Eq:BHZThermalJumpBandEdgeLi3}
\end{align}
\end{widetext}
Eq.~\eqref{Eq:BHZThermalJumpBandEdgeLi3} is the band edge approximation to the finite temperature correction of the domain wall response. The first term gives the leading contribution near $|\mu|\sim |m|$, while the ${\rm Li}_2$ and ${\rm Li}_3$ terms are corrections generated by the subleading terms in the expansion of $\phi(E)$ around $E=|m|$. The physical content is clearest in the leading term. For positive $\mu$ close to the conduction band edge, the second logarithm in $L_1$ is exponentially smaller, and
\begin{align}
    \frac{1}{\beta}L_1(\beta,\mu)
    \simeq
    k_B T\ln\!\left(1+e^{(\mu-|m|)/(k_B T)}\right)\,.
    \label{Eq:BHZThermalJumpCrossoverFunction}
\end{align}
Therefore, finite temperature rounds the sharp zero temperature kink at $|\mu|=|m|$ over an energy window of order $k_BT$. Even when the chemical potential is still inside the gap, thermally occupied states near the band edge modify the domain wall jump by an amount controlled by $(|m|-|\mu|)/(k_BT)$.

The band edge expansion should not be used when the chemical potential is deep in the conduction or valence band. In the regime $|\mu|>|m|$ with $|\mu|-|m|\gg k_BT$, the appropriate approximation is instead the Sommerfeld expansion. Applying the Sommerfeld result on the two sides of the domain wall gives
\begin{widetext}
\begin{align}
    \Delta\delta\zeta_{\rm H}^{\rm S}(\beta,\mu)=
    \Delta\delta\zeta_{\rm H}(T=0,|\mu|) +\frac{\pi b|\mu|}{12\beta^2} \left[ \frac{1}{\left(1-4b|m|+4b^2\mu^2\right)^{3/2}}-\frac{1}{\left(1+4b|m|+4b^2\mu^2\right)^{3/2}}\right] + O(T^4)\,.
    \label{Eq:BHZThermalJumpSommerfeld}
\end{align}
\end{widetext}
Here, $\Delta\delta\zeta_{\rm H}(T=0,|\mu|)$ is the zero temperature finite chemical potential correction to the domain wall jump. Hence, Eq.~\eqref{Eq:BHZThermalJumpBandEdgeLi3} describes the rounded band edge regime $|\mu|\sim |m|$, while Eq.~\eqref{Eq:BHZThermalJumpSommerfeld} describes the regime where the Fermi level is well inside the band. For numerical predictions across the full range of $\mu$, we use the exact integral in Eq.~\eqref{Eq:BHZThermalExactJumpCorrection}.

It is useful to express the result as a prediction for the ratio relative to the pure Dirac domain wall jump,
\begin{align}
    \mathcal R(\beta,\mu)=\frac{\Delta\zeta_{\rm H}(\beta,\mu)}{\Delta\zeta_{\rm H}^{0}}=\mathcal R_0+\frac{\Delta\delta\zeta_{\rm H}(\beta,\mu)}{\Delta\zeta_{\rm H}^{0}}\,.
    \label{Eq:BHZThermalRatioDefinition}
\end{align}
\begin{figure}[!t]
    \centering
    \subfigure[$|\mathcal R(\mu)|$ at fixed $T$.]{
        \includegraphics[scale=0.6]{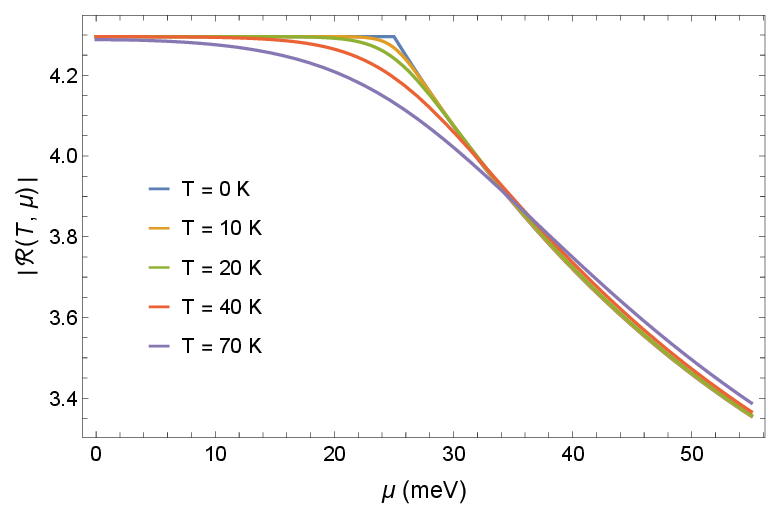}
    }
    \subfigure[$|\mathcal R(T)|$ at fixed $\mu$.]{
        \includegraphics[scale=0.6]{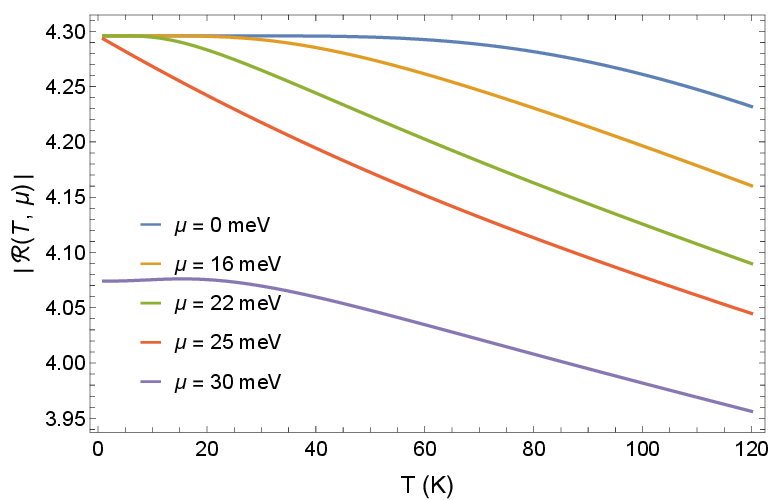}
    }
    \caption{The magnitude of the domain wall ratio $|\mathcal R(T,\mu)|$ for (Hg,Mn)Te parameters. The vacuum value $|\mathcal R_0|\simeq4.3$ is stable when $|m|-|\mu|\gg k_BT$ and is reduced near the thermally rounded band edge.}
    \label{fig:BHZThermalDomainWallRatio}
\end{figure}
Restoring $a=\hbar v_{\rm F}$, the vacuum ratio is given by Eq.~\eqref{Eq:ReviewBHZVacuumRatio}. We have seen that for the same (Hg,Mn)Te parameters, one finds $\mathcal R_0\simeq-4.3$. At finite temperature and chemical potential, the prediction is obtained from Eq.~\eqref{Eq:BHZThermalRatioDefinition}, where $\Delta\delta\zeta_{\rm H}$ is evaluated from the exact integral, from the band edge expansion near $|\mu|\sim |m|$, or from the Sommerfeld expansion for $|\mu|-|m|\gg k_BT$. The resulting behavior is shown in Fig.~\ref{fig:BHZThermalDomainWallRatio}. The ratio $\mathcal R$ is essentially unchanged from its vacuum value $\mathcal R_0\simeq-4.3$ whenever $|m|-|\mu|\gg k_BT$, namely when the chemical potential lies well inside the bulk gap. In this regime the correction is thermally activated as $\exp[-(|m|-|\mu|)/(k_BT)]$. As shown in Fig.~\ref{fig:BHZThermalDomainWallRatio}(a), increasing the temperature broadens the crossover in $\mathcal R(\mu)$, and hence the reduction begins already below the zero temperature band edge $|\mu|=|m|$. Equivalently, Fig.~\ref{fig:BHZThermalDomainWallRatio}(b) shows that, at fixed $|\mu|<|m|$, the magnitude of $\mathcal R(T)$ decreases from that of $\mathcal R_0$ once thermal occupation of band edge states becomes appreciable; the onset occurs at lower temperature when $|\mu|$ is closer to $|m|$. For $|\mu|>|m|$, the zero temperature value is already reduced from $|\mathcal R_0|$, and finite temperature produces the Sommerfeld correction once the chemical potential is well inside the band. The finite temperature and chemical potential correction therefore acts against the vacuum enhancement generated by the BHZ deformation. 

The fixed-$T$ curves in Fig.~\ref{fig:BHZThermalDomainWallRatio}(a) exhibit an approximate clustering at chemical potentials above the band edge. Near the band edge, thermal rounding reduces $|\mathcal R|$, whereas deeper in the band the leading Sommerfeld correction has the opposite sign for $b<0$, causing the ordering of the curves to reverse. At larger $\mu$ or $T$, the differences among the curves decrease, and all curves approach the common continuum limit $\mathcal R_\infty=-a^2/(2|bm|)$, which is approximately $-2.48$ for the (Hg,Mn)Te parameters considered here. However, these extreme chemical potential and temperature limits lie outside the quantitative range of validity of the low-energy BHZ model and should be understood as formal properties of the continuum result.

\section{Conclusions}
\label{sec:Conclusions}
In this work, we have extended the torsional Hall viscosity of Chern insulators to finite temperature and chemical potential. This continues the analysis of Ref.~\cite{matthaiakakis2025}, where the effects of a constant magnetic field and of the momentum-dependent mass appearing in a single block of the BHZ Hamiltonian were studied at zero temperature and zero chemical potential. Our main goal here was to determine how thermal occupation and finite chemical potential modify these vacuum results and, in particular, under which conditions the torsional Hall viscosity remains close to its ground-state value. A general feature of the analysis is that finite $T$ and $\mu$ do not modify the ultraviolet structure of the theory. The same renormalization as in the vacuum can therefore be used, while the corrections due to finite temperature and chemical potential are finite.

For the free massive Dirac fermion, we found that at zero temperature the torsional Hall viscosity remains unchanged as long as the chemical potential lies inside the bulk gap. At finite temperature, the sharp onset associated with crossing a band edge is smoothed by the Fermi--Dirac distribution, while the correction is exponentially suppressed when the chemical potential lies sufficiently deep inside the gap. We then considered a constant magnetic field, for which the response can be organized in terms of the relativistic Landau levels. The zeroth Landau level gives an additional orientation-sensitive contribution at finite chemical potential. At zero temperature, tuning $\mu$ through the higher Landau levels produces a staircase structure in $\delta\zeta_{\rm H}$. In contrast to the Hall conductivity, the step heights are not quantized in a fixed universal unit, but depend on the mass and the Landau-level energy. Moreover, finite temperature rounds the steps over an energy scale of order $k_{\rm B}T$.

We finally studied the momentum-dependent mass of the single-block BHZ model. The correction at finite temperature and chemical potential is ultraviolet finite and can be obtained for arbitrary $T$ and $\mu$ within the continuum model. Away from the band edge, its low-temperature behavior is captured by a Sommerfeld expansion, whereas close to the band edge a separate expansion in terms of polylogarithms describes the thermal rounding of the zero-temperature threshold. We also studied the torsional Hall viscosity jump across a mass domain wall. The enhancement found in Ref.~\cite{matthaiakakis2025} for parameters relevant to (Hg,Mn)Te quantum wells remains essentially unchanged when the chemical potential lies sufficiently deep inside the gap, i.e.,
\begin{align}
    |m|-|\mu|\gg k_{\rm B}T\,,
\end{align}
up to exponentially suppressed corrections. As the chemical potential approaches and enters a bulk band, thermal occupation and finite-density effects reduce the magnitude of the enhanced domain-wall response, while finite temperature smooths the zero-temperature crossover near the band edge.

An important next step is to combine the magnetic-field and momentum-dependent mass deformations. In magnetically doped (Hg,Mn)Te quantum wells, a weak external magnetic field can be used to polarize the Mn moments. Although the role of this field is primarily to align the magnetic moments, its orbital coupling to the electronic states cannot in general be neglected. A direct comparison with this experimental setting therefore calls for the torsional Hall viscosity of the single-block BHZ Hamiltonian in a finite magnetic field, including the resulting Landau-level structure, and eventually at finite temperature and chemical potential. Together with the results obtained here, such an analysis would provide a more complete description of the torsional response under conditions relevant to experiments on topological quantum wells with broken time-reversal symmetry.

\begin{acknowledgments}

We thank Christian Tutschku and Raffael Klees for collaboration in the early stages of this project. W.J.~thanks Jingyuan Chen, Zheng-Cheng Gu and Ze-Min Huang for helpful discussions. E.M.H.~thanks Joel Moore and James Analytis for the insightful discussions and the support as well as the hospitality at the University of California, Berkeley. E.M.H., J.E., and R.M.~acknowledge the support of the German Research Foundation (DFG) through the Collaborative Research Center ToCoTronics, Project-ID 258499086 — SFB 1170, as well as Germany’s Excellence Strategy through the W{\"u}rzburg-Dresden Cluster of Excellence on Complexity, Topology and Dynamics in Quantum Matter - ctd.qmat (EXC 2147, Project-ID 390858490). R.M. furthermore acknowledges hospitality from the Shanghai Institute
for Mathematics and Interdisciplinary Sciences (SIMIS) and associated travel support under
STCSM Grant 25HB2701900. W.J.~is supported by the Research Grants Council of Hong Kong Special Administrative Region of China (Project No.~14302725) under the scheme of General Research Fund. The work of I.M.~is supported by the STFC consolidated grant (ST/X000583/1) `New Frontiers In Particle Physics, Cosmology And Gravity'. Z.Y.X. acknowledges support from the basic startup funding from Sun Yat-sen University and from the Berlin Quantum Initiative.
\end{acknowledgments}

\appendix

\onecolumngrid

\section{Evaluating $\zeta_{\rm H}$ at finite temperature and chemical potential}\label{App:MatsSum}

In the present appendix, we show how to evaluate the torsional Hall viscosity $\zeta_{\rm H}$ at finite temperature and chemical potential. To begin our calculation, recall that $\zeta_{\rm H}$ is given by Eq.~\eqref{Eq:ZetaMats} as 
\begin{equation}
    \label{AppEq:ZetaMats}
    \zeta_{\rm H}(\beta,\mu;\Lambda)
    = {m\over 4\pi}\left[S(\beta,\mu;m_\Lambda)-S(\beta,\mu;|m|)\right]\,,
\end{equation}
where $m_\Lambda=\sqrt{\Lambda^2+m^2}$, and we have defined the Matsubara sum $S(\beta,\mu;x)$ as
\begin{equation}
\label{AppEq:MatsSumDef}
    S(\beta,\mu;x) = {1\over \beta}\sum_{\omega_l}\ln\left[{(\omega_l +i\mu)^2 +x^2}\right]\,.
\end{equation}
The individual sums in Eq.~\eqref{AppEq:MatsSumDef} are formal; their difference in Eq.~\eqref{AppEq:ZetaMats} converges at fixed $\Lambda$. We evaluate this difference before taking $\Lambda\rightarrow\infty$. To evaluate $S(\beta,\mu;m_\Lambda)-S(\beta,\mu;|m|)$, we shall employ complex analysis. To this end, we introduce two auxiliary functions
\begin{equation}
    \label{AppEq:AuxF}
    n(z) = {1 \over 1+\exp\left[i\beta(z-i\mu)\right]}\,,\qquad N(z) = \ln\left[1 + \exp\left(-i\beta(z-i\mu)\right)\right]\,.
\end{equation}
The two functions $n(z)$ and $N(z)$ are not independent, but related as $n(z) = i\partial_z N(z)/\beta$. Importantly, $n(z)$ has poles at $z=\omega_l+i\mu$ with residue
\begin{equation}
    \label{AppEq:Residues}
    {\rm Res}[n(z\rightarrow \omega_l +i\mu)] = {i\over \beta}\,.
\end{equation}
The function $N(z)$ has branch points at the poles of $n(z)$.

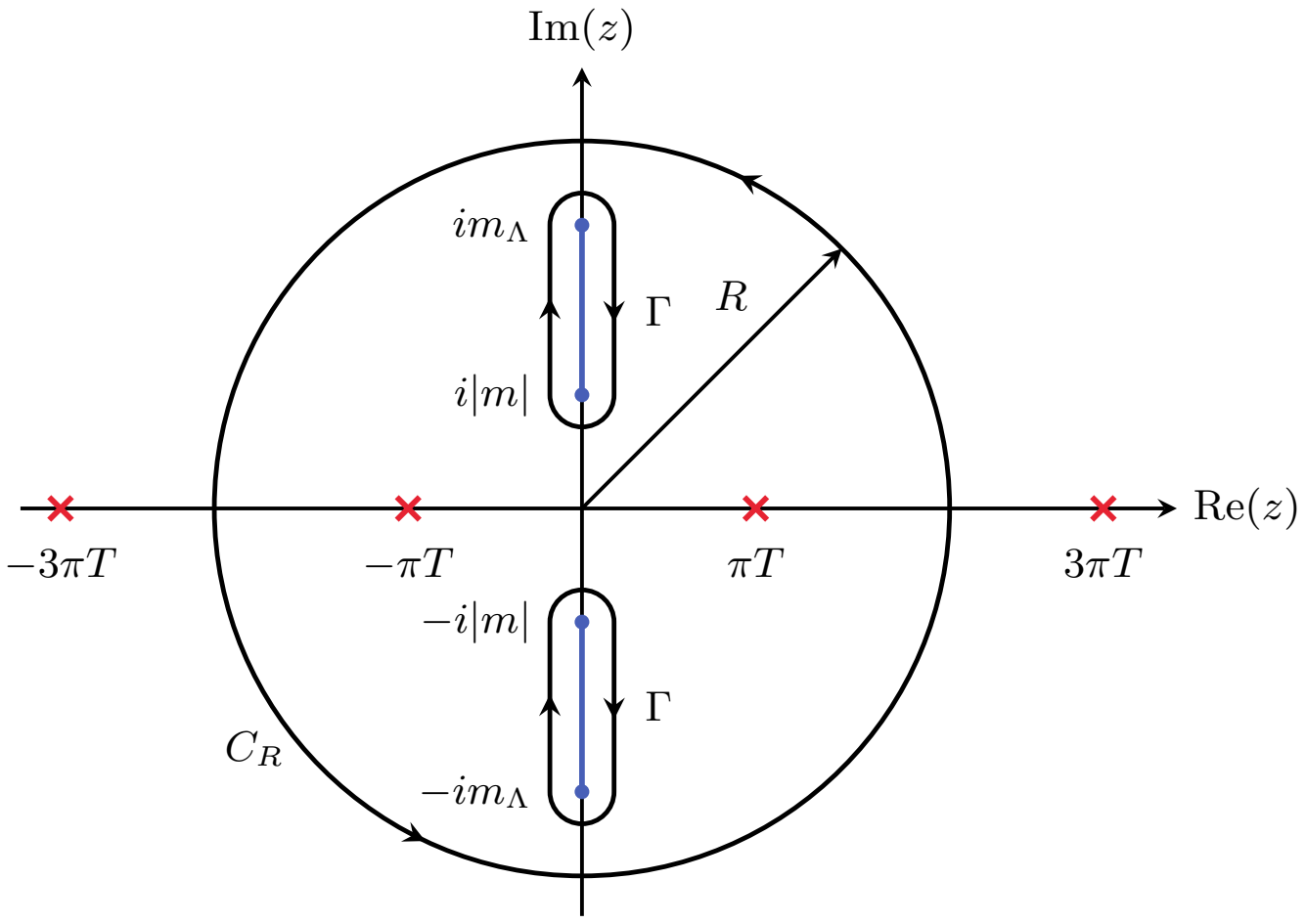
\begin{figure}[h]
    \centering
    \resizebox{0.5\linewidth}{!}{%
    \begin{tikzpicture}[x=1cm,y=1cm,>=stealth,line width=0.8pt,
      contour/.style={black,line width=1pt,postaction={decorate},
        decoration={markings,mark=at position 0.18 with {\arrow{>}},
          mark=at position 0.68 with {\arrow{>}}}}]
      \definecolor{cutblue}{RGB}{72,95,183}
      \definecolor{polered}{RGB}{230,35,50}
      \draw[->,black] (-4.2,0)--(4.45,0) node[right] {$\text{Re}(z)$};
      \draw[->,black] (0,-3.05)--(0,3.3) node[above] {$\text{Im}(z)$};
      \draw[contour] (2.75,0) arc[start angle=0,end angle=360,radius=2.75];
      \draw[->,black] (0,0)--(1.945,1.945) node[pos=0.7,above left] {$R$};
      \node[black] at (-2.45,-1.8) {$C_R$};
      \foreach \x/\lab in {-3.9/{-3\pi T},-1.3/{-\pi T},1.3/{\pi T},3.9/{3\pi T}} {
        \draw[polered,line width=1.2pt] (\x-0.085,-0.085)--(\x+0.085,0.085)
          (\x-0.085,0.085)--(\x+0.085,-0.085);
        \node[below=5pt,black] at (\x,0) {$\lab$};
      }
      \draw[cutblue,line width=1.2pt] (0,0.85)--(0,2.12);
      \draw[cutblue,line width=1.2pt] (0,-2.12)--(0,-0.85);
      \foreach \y/\lab in {0.85/{i|m|},2.12/{im_\Lambda},-0.85/{-i|m|},-2.12/{-im_\Lambda}} {
        \fill[cutblue] (0,\y) circle (0.055);
        \node[anchor=east,inner sep=1pt,black] at (-0.33,\y) {$\lab$};
      }
      \foreach \lo/\hi in {0.85/2.12,-2.12/-0.85} {
        \draw[contour] (0.24,\hi)--(0.24,\lo)
          arc[start angle=0,end angle=-180,radius=0.24]
          --(-0.24,\hi) arc[start angle=180,end angle=0,radius=0.24]--cycle;
      }
      \node[right,black] at (0.35,1.48) {$\Gamma$};
      \node[right,black] at (0.35,-1.48) {$\Gamma$};
    \end{tikzpicture}%
    }
    \caption{Contour $C=C_R+\Gamma$, with an outer circle of radius $R$, for performing the integral ${\cal I}$ in Eq.~\eqref{AppEq:Integral1}, for simplicity shown at $\mu =0$. The crosses designate the poles of the integral along the real axis, while the small disks mark the four branch points along the imaginary axis. The two cuts connect $i|m|$ to $im_\Lambda$ and $-im_\Lambda$ to $-i|m|$.}
    \label{Fig:Contour1}
\end{figure}

Consider now the integral
\begin{equation}
\label{AppEq:Integral1}
    {\cal I} = -i\int_C dz\,L(z)n(z)\,,\qquad
    L(z)=\ln\left[\frac{z^2+m_\Lambda^2}{z^2+m^2}\right]\,,
\end{equation}
where $C$ is the contour of Fig.~\ref{Fig:Contour1}. We choose the branch of $L(z)$ that vanishes at infinity, with cuts from $i|m|$ to $im_\Lambda$ and from $-im_\Lambda$ to $-i|m|$. The contour $C=C_R+\Gamma$ consists of an outer counterclockwise circle $C_R$ and two clockwise contours $\Gamma$ enclosing the excluded cuts.

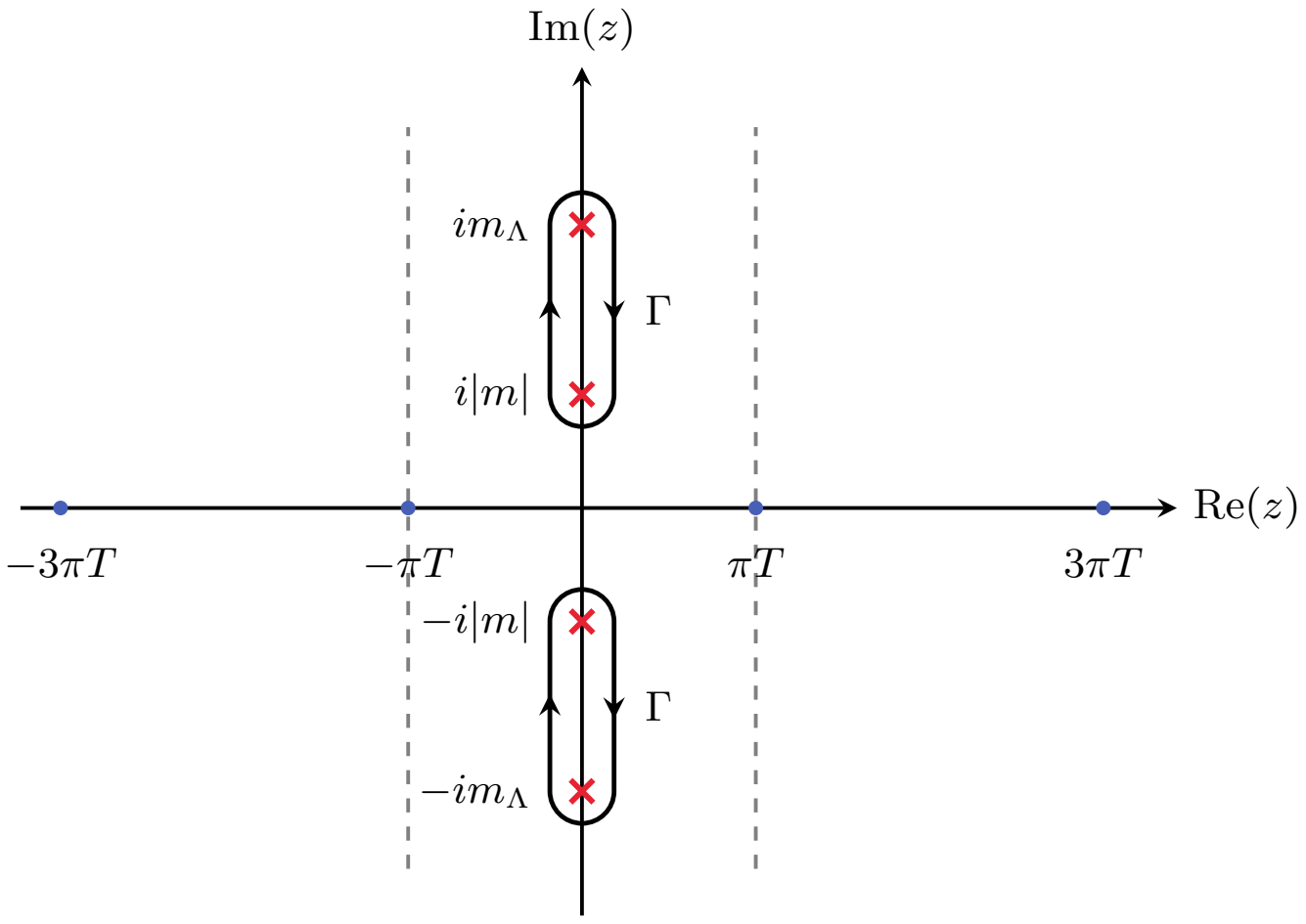
\begin{figure}[h]
    \centering
    \resizebox{0.5\linewidth}{!}{%
    \begin{tikzpicture}[x=1cm,y=1cm,>=stealth,line width=0.8pt,
      contour/.style={black,line width=1pt,postaction={decorate},
        decoration={markings,mark=at position 0.18 with {\arrow{>}},
          mark=at position 0.68 with {\arrow{>}}}}]
      \definecolor{cutblue}{RGB}{72,95,183}
      \definecolor{polered}{RGB}{230,35,50}
      \draw[gray,dashed] (-1.3,-2.7)--(-1.3,2.85);
      \draw[gray,dashed] (1.3,-2.7)--(1.3,2.85);
      \draw[->,black] (-4.2,0)--(4.45,0) node[right] {$\text{Re}(z)$};
      \draw[->,black] (0,-3.05)--(0,3.3) node[above] {$\text{Im}(z)$};
      \foreach \x/\lab in {-3.9/{-3\pi T},-1.3/{-\pi T},1.3/{\pi T},3.9/{3\pi T}} {
        \fill[cutblue] (\x,0) circle (0.055);
        \node[below=5pt,black] at (\x,0) {$\lab$};
      }
      \foreach \y/\lab in {0.85/{i|m|},2.12/{im_\Lambda},-0.85/{-i|m|},-2.12/{-im_\Lambda}} {
        \draw[polered,line width=1.2pt] (-0.085,\y-0.085)--(0.085,\y+0.085)
          (-0.085,\y+0.085)--(0.085,\y-0.085);
        \node[anchor=east,inner sep=1pt,black] at (-0.33,\y) {$\lab$};
      }
      \foreach \lo/\hi in {0.85/2.12,-2.12/-0.85} {
        \draw[contour] (0.24,\hi)--(0.24,\lo)
          arc[start angle=0,end angle=-180,radius=0.24]
          --(-0.24,\hi) arc[start angle=180,end angle=0,radius=0.24]--cycle;
      }
      \node[right,black] at (0.35,1.48) {$\Gamma$};
      \node[right,black] at (0.35,-1.48) {$\Gamma$};
    \end{tikzpicture}%
    }
    \caption{Clockwise contours $\Gamma$ for performing the integral ${\cal I}$ in Eq.~\eqref{AppEq:AuxI1}, for simplicity shown at $\mu =0$. The crosses designate the four poles at $z=\pm i|m|,\pm im_\Lambda$, while the small disks mark the branch points of $N(z)$ along the real axis. Both contours lie inside the dashed lines, where $N(z)$ is analytic.}
    \label{Fig:Contour2}
\end{figure}

With this choice of contour, we encircle the poles of $n(z)$ inside $C_R$ and do not cross either branch cut. We can then employ the residue theorem to evaluate ${\cal I}$, and find in the limit $R\rightarrow\infty$
\begin{equation}
    \label{AppEq:S=Mats}
    {\cal I} = 2\pi i\left[S(\beta,\mu;m_\Lambda)-S(\beta,\mu;|m|)\right]\,.
\end{equation}
This establishes that if we calculate ${\cal I}$, we can calculate the difference in Eq.~\eqref{AppEq:ZetaMats}. At fixed $\Lambda$, $L(z)=O(z^{-2})$. Taking $R\rightarrow\infty$ through circles that stay a fixed distance from the Matsubara poles, $n(z)$ remains bounded and the outer-circle integral vanishes as $O(R^{-1})$. Hence, only the clockwise contours $\Gamma$ contribute. To calculate ${\cal I}$ directly, we integrate by parts to introduce $N(z)$ under the integral as
\begin{equation}
    \label{AppEq:AuxI1}
    \begin{aligned}
    {\cal I} &= {1\over \beta}\int_{\Gamma} d N(z)\,L(z)
    \\
    &= -{2 \over \beta}\int_{\Gamma} dz~\left[\frac{z}{z^2+m_\Lambda^2}-\frac{z}{z^2+m^2}\right]N(z)\,.
    \end{aligned}
\end{equation}
The contours $\Gamma$ shown in Fig.~\ref{Fig:Contour2} lie within $|\text{Re}z|<\pi/\beta$, where $N(z)$ has an analytic branch. Both $L(z)$ and $N(z)$ return to their starting values along each closed contour, and hence $\int_\Gamma d[N(z)L(z)]=0$ and the integration-by-parts boundary term vanishes.
The resulting integrand has poles only at $z=\pm i|m|,\pm im_\Lambda$ inside $\Gamma$. Therefore, we can once again use Cauchy's theorem with clockwise orientation to find
\begin{equation}
\begin{aligned}
    {\cal I} &= -\frac{2}{\beta}(-2\pi i)\left[\frac{N(im_\Lambda)+N(-im_\Lambda)}{2}-\frac{N(i|m|)+N(-i|m|)}{2}\right]
    \\
    &=\frac{2\pi i}{\beta}\left[N(im_\Lambda)+N(-im_\Lambda)-N(i|m|)-N(-i|m|)\right]
    \\
    &=\frac{2\pi i}{\beta}\ln\left[\frac{\cosh\left(\beta(m_\Lambda-\mu)/2\right)\cosh\left(\beta(m_\Lambda+\mu)/2\right)}{\cosh\left(\beta(|m|-\mu)/2\right)\cosh\left(\beta(|m|+\mu)/2\right)}\right]\,,
\end{aligned}
\end{equation}
which implies for $\zeta_{\rm H}(\beta,\mu;\Lambda)$
\begin{equation}
    \label{AppEq:MatsSumResult}
    \zeta_{\rm H}(\beta,\mu;\Lambda)=\frac{m}{4\pi\beta}\ln\left[\frac{\cosh\left(\beta(m_\Lambda-\mu)/2\right)\cosh\left(\beta(m_\Lambda+\mu)/2\right)}{\cosh\left(\beta(|m|-\mu)/2\right)\cosh\left(\beta(|m|+\mu)/2\right)}\right]\,.
\end{equation}

Equation \eqref{AppEq:MatsSumResult} is our final result for the regulated torsional Hall viscosity. At fixed $\beta$ and $\mu$, the logarithm of the numerator is $\beta m_\Lambda-\ln4+O(e^{-\beta(m_\Lambda-|\mu|)})$. Using $m_\Lambda=\Lambda+O(\Lambda^{-1})$ and taking $\Lambda\rightarrow\infty$ reproduces Eq.~\eqref{Eq:ThermalZeta} presented in the main text. The vacuum counterterms then give the renormalized response in Eq.~\eqref{Eq:DiracThermalRenormalizedFinitePart}.

\section{Laplace transform of $\zeta_{\rm H}$ at finite magnetic field}
\label{App:Laplace}

We evaluate the differential operator in Eq.~\eqref{Eq:ZetaTBM} spectrally. Set $x=m^2$, $a=l_B^{-2}$, and $D=\partial_x$, and define
\begin{align}
\label{AppEq:LaplaceOperatorDefinition}
{\cal A}_a&=aD\coth(aD)\,,\nn\\
f(x)&=\frac{\delta\zeta_{\rm H}(\beta,\mu;m)}{m}=-\frac{1}{4\pi\beta}\left\{
\ln\left[1+e^{-\beta(\sqrt{x}-\mu)}\right]
+\ln\left[1+e^{-\beta(\sqrt{x}+\mu)}\right]
\right\}\,.
\end{align}
On either gapped branch the sign of the Dirac mass is carried by the overall factor of $m$, whereas $f$ depends only on $x=m^2$. For $x>0$, the function $f$ is smooth and decays exponentially in $\sqrt{x}$ as $x\rightarrow\infty$. We first take $\beta>0$ and $x>\mu^2$, where the logarithms admit convergent expansions in $e^{-\beta(\sqrt{x}\mp\mu)}$. Using the identity, for $c>0$,
\begin{align*}
e^{-c\sqrt{x}}=\frac{c}{2\sqrt{\pi}}\int_0^\infty d\tau\,\tau^{-3/2}e^{-c^2/(4\tau)}e^{-\tau x}\,,
\end{align*}
we obtain the Laplace representation
\begin{align}
\label{AppEq:SpectralLaplaceRepresentation}
f(x)=\int_0^\infty d\tau\,\rho(\tau)e^{-\tau x}\,,
\end{align}
with
\begin{align*}
\rho(\tau)=-\frac{1}{4\pi^{3/2}\tau^{3/2}}\sum_{r=1}^{\infty}(-1)^{r+1}\cosh(r\beta\mu)e^{-r^2\beta^2/(4\tau)}\,.
\end{align*}
For $x>\mu^2$, these expansions and the integrals below converge absolutely, justifying the interchanges of sums, integrals, and derivatives.
The spectral weight $\rho(\tau)$ need not be positive. This representation diagonalizes $D$. Since $D e^{-\tau x}=-\tau e^{-\tau x}$ and $z\coth z$ is even, one obtains
\begin{align}
\label{AppEq:DifferentialOperatorSpectral}
{\cal A}_a f(x)
=\int_0^\infty d\tau\,\rho(\tau)
a\tau\coth(a\tau)e^{-\tau x}\,.
\end{align}
For $a\tau>0$,
\begin{align}
\label{AppEq:CothConvergentExpansion}
\coth(a\tau)=1+2\sum_{n=1}^{\infty}e^{-2na\tau}\,.
\end{align}
Substituting Eq.~\eqref{AppEq:CothConvergentExpansion} into Eq.~\eqref{AppEq:DifferentialOperatorSpectral} and differentiating Eq.~\eqref{AppEq:SpectralLaplaceRepresentation} gives
\begin{align}
\label{AppEq:CorrectedLaplaceResult}
{\cal A}_a f(x)
=-a\left[f'(x)+2\sum_{n=1}^{\infty}f'(x+2na)\right]\,.
\end{align}
For the function in Eq.~\eqref{AppEq:LaplaceOperatorDefinition},
\begin{align}
\label{AppEq:SpectralThermalDerivative}
f'(x)
=\frac{1}{8\pi\sqrt{x}}
\left[
n_{\rm F}(\sqrt{x}-\mu)
+n_{\rm F}(\sqrt{x}+\mu)
\right]\,.
\end{align}
Writing
\begin{align}
E_n=\sqrt{x+2na}\,,\qquad
g_n=2-\delta_{n0}\,,\qquad n=0,1,\ldots\,,
\end{align}
Eq.~\eqref{AppEq:CorrectedLaplaceResult} becomes
\begin{align}
\label{AppEq:SpectralLandauSum}
{\cal A}_a f(x)
=-\frac{1}{8\pi l_B^2}
\sum_{n=0}^{\infty}\frac{g_n}{E_n}
\left[
n_{\rm F}(E_n-\mu)+n_{\rm F}(E_n+\mu)
\right]\,.
\end{align}
Multiplication by the overall factor of $m$ in Eq.~\eqref{Eq:ZetaTBM} reproduces Eq.~\eqref{Eq:ZetaBDifferentialLandauSum}. At fixed $x>0$ and $\beta>0$, the Landau-level sum converges locally uniformly and is analytic in $\mu$ throughout $|\text{Im}\mu|<\pi/\beta$. It therefore continues the mass-trace result from $|\mu|<\sqrt{x}$ to all real $\mu$; the Laplace integral itself need not converge outside its initial domain. The zero-temperature result is obtained by taking $\beta\rightarrow\infty$ afterwards. The continued Landau-level expression defines the differential operator for $m^2>0$ and requires no boundary data at $m^2=0$. It gives the complete mass-trace correction at fixed nonzero mass; the massless point, when needed, is understood as a one-sided limit of the gapped response.

\section{Validity of the BHZ low-temperature expansions}
\label{app:BHZValidity}

In this appendix we estimate the regimes of validity of the two analytic approximations used in Sec.~\ref{sec:BHZThermal}.

\begin{figure}[h]
    \centering
    \subfigure[]{\includegraphics[scale=0.6]{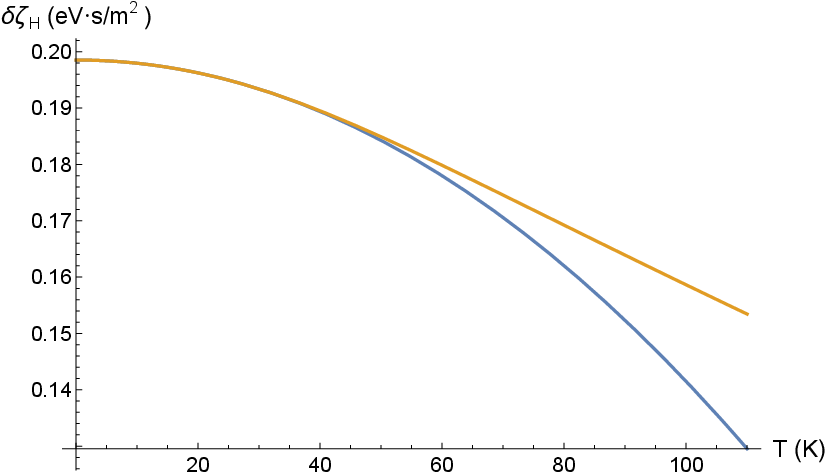}}\quad
    \subfigure[]{\includegraphics[scale=0.6]{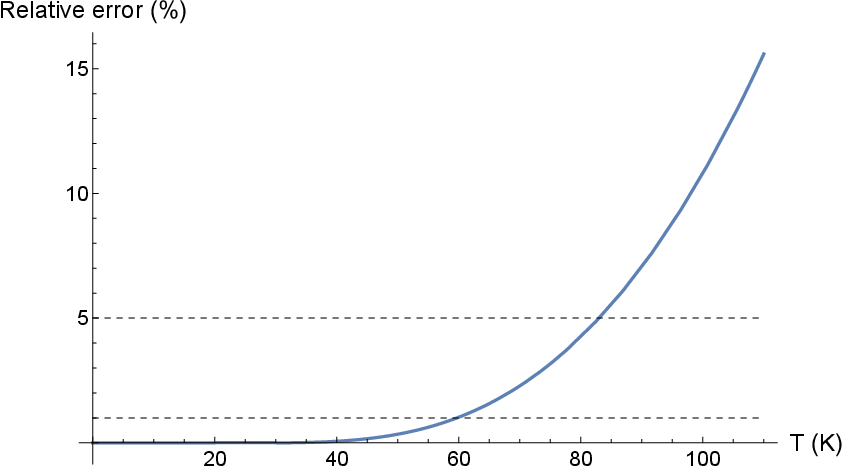}}
    \caption{(a) Numerical (orange) and Sommerfeld approximation (blue) results for $\delta\zeta_H$ as a function of $T$ at $\mu=45$ meV. (b) The error of Sommerfeld approximation compared to the numerical result, with 1\% and 5\% dashed lines.}
    \label{fig:BHZSerror}
\end{figure}

First consider the Sommerfeld expansion for $\mu>|m|$. After changing variables from $k$ to $E$, the finite-temperature correction contains an integral of the form
\begin{align}
    \int_{|m|}^{\infty} dE\,\phi(E)n_{\rm F}(E-\mu)\,.
\end{align}
At $T=0$, $n_{\rm F}(E-\mu)$ becomes a step function at $E=\mu$. At finite temperature this step is rounded over an interval of order $k_BT$. The Sommerfeld expansion assumes that the lower endpoint $E=|m|$ lies outside this interval, and hence the endpoint correction is exponentially small, with leading behavior
\begin{align}
    \exp\!\left[-\frac{\mu-|m|}{k_BT}\right]\,.
\end{align}
A useful estimate of the Sommerfeld validity scale is obtained by choosing a small $\epsilon$ and defining $T_{\rm S}(\epsilon)$ by
\begin{align}
    \exp\!\left[-\frac{\mu-|m|}{k_BT_{\rm S}}\right]=\epsilon\,,
\end{align}
which gives
\begin{align}
    k_BT_{\rm S}(\epsilon)=\frac{\mu-|m|}{\ln(1/\epsilon)}\,.
    \label{Eq:BHZValiditySommerfeldScale}
\end{align}
The Sommerfeld expansion is then expected to be quantitatively reliable for $T\lesssim T_{\rm S}$, provided the omitted higher-order terms are also small. For $1\%$ and $5\%$ criteria, we have
\begin{align}
    k_BT_{\rm S}(1\%)\simeq\frac{\mu-|m|}{4.61}\,,\qquad
    k_BT_{\rm S}(5\%)\simeq\frac{\mu-|m|}{3.00}\,.
\end{align}
For example, for (Hg,Mn)Te we have $|m|=25~{\rm meV}$ and we take $\mu=45~{\rm meV}$, corresponding to $\mu-|m|=20~{\rm meV}$, this gives
\begin{align}
    T_{\rm S}(1\%)\simeq 50~{\rm K}\,,\qquad
    T_{\rm S}(5\%)\simeq 77~{\rm K}\,.
\end{align}
It is useful to compare these analytic estimates with the direct numerical evaluation of the full finite-temperature integral. The Sommerfeld approximation up to $O(T^2)$ differs from the numerical result by about $1\%$ at $T\simeq59$~K and by about $5\%$ at $T\simeq 82$~K (see Fig.~\ref{fig:BHZSerror}). Hence, the above estimation is consistent with the comparison between the Sommerfeld approximation and the full numerical result.

We now estimate the regime of validity of the band-edge expansion used for the finite-temperature BHZ correction. The band-edge expansion is obtained by Taylor expanding $\phi(E)$ about $E=|m|$,
\begin{align}
    \phi(E)=\phi(|m|)+\phi'(|m|)(E-|m|)+\frac{1}{2}\phi''(|m|)(E-|m|)^2+\frac{1}{6}\phi'''(|m|)(E-|m|)^3+\cdots\,,
\end{align}
and then performing the Fermi integrals exactly. This expansion also requires the chemical potential to remain sufficiently close to the band edge for the Taylor expansion of $\phi(E)$ to be accurate. In the regime of positive $\mu$ near the conduction-band edge, the antiparticle contribution $n_{\rm F}(E+\mu)$ is exponentially suppressed, and the expansion of $\delta I=-2\pi\delta\zeta_{\rm H}$ takes the form
\begin{align}
    \delta I(\beta,\mu)=A_1 k_B T\,L_1(x)+A_2 (k_B T)^2 L_2(x)+A_3 (k_B T)^3 L_3(x)+A_4 (k_B T)^4 L_4(x)+\cdots\,,
    \label{Eq:AppBandEdgeSeries}
\end{align}
where
\begin{align}
    x=\frac{\mu-|m|}{k_B T}\,,\qquad L_1(x)=\ln(1+e^x)\,,\qquad L_n(x)={\rm Li}_n(-e^x)\,,\qquad n\ge 2\,.
    \label{Eq:AppBandEdgeLfunctions}
\end{align}
The first four coefficients are
\begin{align}
    A_1&=\frac{m}{2(1-2bm)}\,,\qquad A_2=\frac{b|m|}{(1-2bm)^3}\,,\\
    A_3&=-\frac{b(-1+4bm+8b^2m^2)}{(1-2bm)^5}\,,\qquad
    A_4=-\frac{12b^3|m|\left(3-12bm-8b^2m^2\right)}{(1-2bm)^7}\,.
    \label{Eq:AppBandEdgeA4}
\end{align}

A simple estimate of the truncation error is obtained by comparing the first omitted term with the retained result. If the expansion is kept through ${\rm Li}_2$, the first omitted contribution is the ${\rm Li}_3$ term, and we define
\begin{align}
    \epsilon_2^{\rm BE}(T,\mu)=\left|\frac{A_3 (k_B T)^3 L_3(x)}{A_1 k_B T\,L_1(x)+A_2 (k_B T)^2 L_2(x)}\right|\,.
    \label{Eq:AppBandEdgeErrorLi2}
\end{align}
Similarly, if the expansion is kept through ${\rm Li}_3$, the first omitted contribution is the ${\rm Li}_4$ term, and we define
\begin{align}
    \epsilon_3^{\rm BE}(T,\mu)=\left|\frac{A_4 (k_B T)^4 L_4(x)}{A_1 k_B T\,L_1(x)+A_2 (k_B T)^2 L_2(x)+A_3 (k_B T)^3 L_3(x)}\right|\,.
    \label{Eq:AppBandEdgeErrorLi3}
\end{align}
In analogy with the Sommerfeld case, we define the band-edge validity scale $T_{\rm BE}^{(N)}(\epsilon)$ as the temperature at which the first omitted term estimate for the truncation after ${\rm Li}_N$ reaches a prescribed tolerance $\epsilon_N^{\rm BE}(T_{\rm BE}^{(N)},\mu)$. Thus, $T_{\rm BE}^{(2)}$ is determined from Eq.~\eqref{Eq:AppBandEdgeErrorLi2}, while $T_{\rm BE}^{(3)}$ is determined from Eq.~\eqref{Eq:AppBandEdgeErrorLi3}.

For (Hg,Mn)Te parameters we take
\begin{align}
    m=-25~{\rm meV}\,,\qquad b=-1075~{\rm meV\,nm^2}\,,\qquad a=365~{\rm meV\,nm}\,.
    \label{Eq:AppHgMnTeParams}
\end{align}
To apply the above formulas to these physical units, one replaces $b\to b/a^2$ in the preceding expressions, where $a$ had been set to $1$. Taking $\mu=30~{\rm meV}$, for which $\mu-|m|=5~{\rm meV}$, the estimator temperatures are
\begin{align}
    T_{\rm BE}^{(2)}(1\%)\simeq26~{\rm K}\,,\qquad T_{\rm BE}^{(2)}(5\%)\simeq64~{\rm K}\,,
    \label{Eq:AppBandEdgeLi2Estimate}
\end{align}
and
\begin{align}
    T_{\rm BE}^{(3)}(1\%)\simeq43~{\rm K}\,,\qquad T_{\rm BE}^{(3)}(5\%)\simeq75~{\rm K}\,.
    \label{Eq:AppBandEdgeLi3Estimate}
\end{align}

\begin{figure}[h!]
    \centering
    \subfigure[]{\includegraphics[scale=0.6]{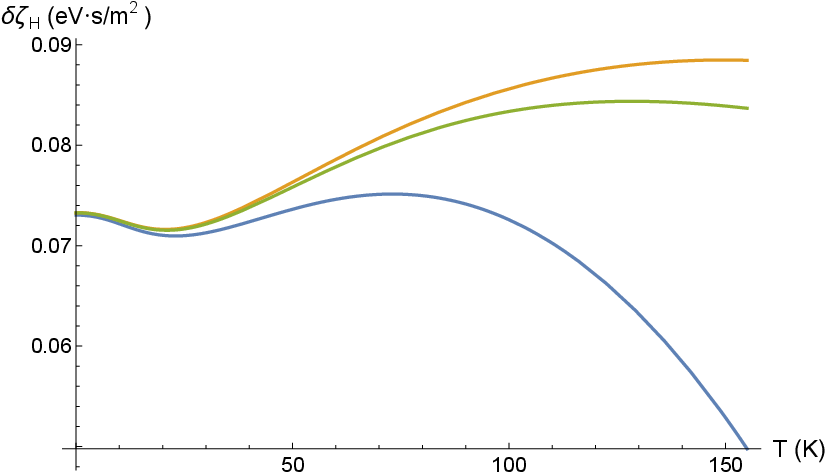}}\quad
    \subfigure[]{\includegraphics[scale=0.6]{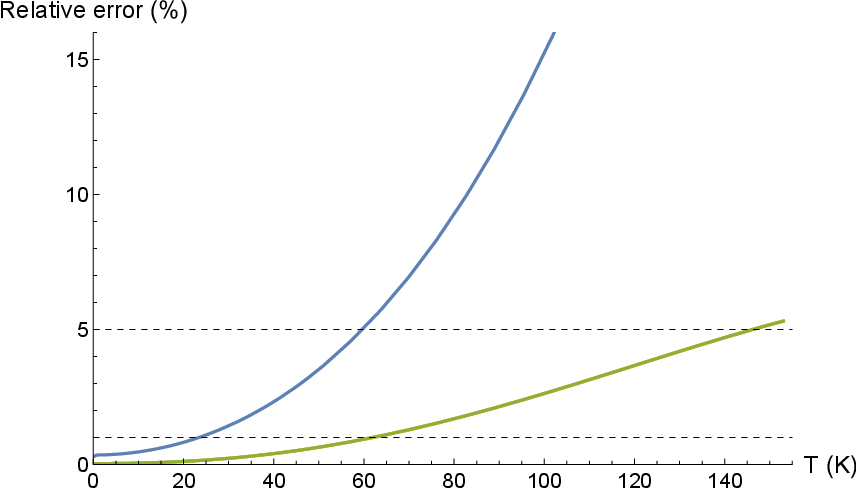}}
    \caption{(a) Numerical result (orange) and the band edge approximations truncated at ${\rm Li}_2$ (blue) and ${\rm Li}_3$ (green) for $\delta\zeta_H$ as a function of $T$ at $\mu=30$ meV. (b) The error of the band edge approximations truncated at ${\rm Li}_2$ (blue) and ${\rm Li}_3$ (green) compared to the numerical result, with 1\% and 5\% dashed lines.}
    \label{fig:BHZBDerror}
\end{figure}

Now we compare these analytic estimates with the direct numerical evaluation. For $\mu=30~{\rm meV}$, the approximation truncated at ${\rm Li}_2$ differs from the numerical result by about $1\%$ at $T\simeq24~{\rm K}$ and by about $5\%$ at $T\simeq62~{\rm K}$. Including the ${\rm Li}_3$ term substantially improves the agreement: the corresponding differences are about $1\%$ at $T\simeq64~{\rm K}$ and about $5\%$ at $T\simeq145~{\rm K}$ (see Fig.~\ref{fig:BHZBDerror}). Thus, the first omitted term estimate captures the correct temperature scale for the breakdown of the ${\rm Li}_2$ approximation and gives a reasonable theoretical estimate for the onset of deviations in the ${\rm Li}_3$ approximation. The actual quantitative accuracy, however, is best determined by direct comparison with the numerical integral. For QAH experiments, which are typically performed at temperatures well below these scales, the band-edge expansion is well controlled in the experimentally relevant regime.

\bibliography{refs.bib}

\end{document}